\documentclass[fleqn,usenatbib]{mnras}
\usepackage{newtxtext,newtxmath}
\usepackage[T1]{fontenc}

\DeclareRobustCommand{\VAN}[3]{#2}
\let\VANthebibliography\thebibliography
\def\thebibliography{\DeclareRobustCommand{\VAN}[3]{##3}\VANthebibliography}

\usepackage{graphicx}	
\usepackage{amsmath}	
\usepackage[dvipsnames]{xcolor}
\usepackage{enumitem}
\usepackage{url}
\usepackage{braket}
\usepackage{threeparttable}
\usepackage{hyperref}
\usepackage{multirow, tabularx}

\def\softwarenamestyle[#1]{\textsc{#1}}
\def\figbf[#1]{{\bf #1}}

\setlist[itemize,1]{label=$\bullet$}
\setlist[itemize,2]{label=$\bullet$}
\setlist[itemize,3]{label=$\bullet$}
\setlist[itemize,4]{label=$\bullet$}
\setlist[itemize]{leftmargin=*}
\def\bit{\begin{itemize}[topsep=0em,parsep=0em,itemsep=0em,partopsep=0em,leftmargin=*]}
\def\bitt{\begin{itemize}[topsep=0em,parsep=0em,itemsep=0em,partopsep=0em,leftmargin=3.0em]}
\def\eit{\end{itemize}}
\def\benum{\begin{enumerate}[leftmargin=1em,itemsep=0pt,parsep=0pt,topsep=0pt]}
\def\eenum{\end{enumerate}}

\def\beq{\begin{equation}}
\def\eeq{\end{equation}}
\def\bey{\begin{eqnarray}}
\def\eey{\end{eqnarray}}

\def\mpc{\, h^{-1}{\rm {Mpc}}}

\def\Msun{{\rm M_\odot}}
\def\msun{\, h^{-1}{\rm M_\odot}}

\def\Mhalo{M_{\rm halo}}

\title[]{Massive Dark Matter Halos at High Redshift: Implications for Observations in the JWST Era}

\author[Yangyao Chen et al.]{
    Yangyao Chen,$^{1,2}$\thanks{E-mail: yangyaochen.astro@foxmail.com}  
    H.J. Mo, $^{3}$   		
    Kai Wang $^{4}$		    
    \\ 
    $^{1}$School of Astronomy and Space Science, University of Science and Technology of China, Hefei, Anhui 230026, China\\
    $^{2}$Key Laboratory for Research in Galaxies and Cosmology, Department of Astronomy, University of Science and Technology of China, Hefei, Anhui 230026, China \\
    $^{3}$Department of Astronomy, University of Massachusetts, Amherst, MA 01003-9305, USA\\
    $^{4}$Kavli Institute for Astronomy and Astrophysics, Peking University, Beijing 100871, China\\
}

\date{Accepted XXX. Received YYY; in original form ZZZ}

\pubyear{2021}

\begin{document}
\label{firstpage}
\pagerange{\pageref{firstpage}--\pageref{lastpage}}
\maketitle


\begin{abstract}
The presence of massive galaxies at high $z$ as recently observed by JWST 
appears to contradict the current $\Lambda$CDM cosmology. Here we aim to alleviate 
this tension by incorporating uncertainties from three sources in counting galaxies: 
cosmic variance, error in stellar mass estimation, and backsplash 
enhancement. Each of these factors significantly increases the cumulative stellar 
mass density $\rho_*(>M_*)$ at the high-mass end, and their combined effect can 
boost the density by more than one order of magnitude. Assuming a star formation 
efficiency of $\epsilon_* \sim 0.5$, cosmic variance alone reduces the tension to 
a $2\sigma$ level, except for the most massive galaxy at $z=8$. Additionally, 
incorporating a 0.3 dex lognormal dispersion in the stellar mass estimation 
brings the observed $\rho_*(>M_*)$ at $z \sim 7 - 10$ within $2\sigma$. The tension 
is completely eliminated when we account for the gas stripped from backsplash 
halos. These results highlight the importance of fully modeling uncertainties when 
interpreting observational data of rare objects. We use the constrained simulation, 
ELUCID, to investigate the descendants of high-$z$ massive galaxies. Our findings 
reveal that a significant portion of these galaxies ultimately reside in massive 
halos at $z=0$ with $M_{\rm halo} > 10^{13}\msun$. Moreover, a large fraction of 
local central galaxies in $M_{\rm halo} \geqslant 10^{14.5} \msun$ halos  are predicted 
to contain substantial amounts of ancient stars formed in massive galaxies at 
$z \sim 8$. This prediction can be tested by studying the structure and 
stellar population of central galaxies in present-day massive 
clusters.
\end{abstract}
\begin{keywords}
galaxies: haloes - galaxies: formation - galaxies: evolution - galaxies: high-redshift - galaxies: statistics
\end{keywords}

\section{Introduction}
\label{sec:intro}

The launch of the James Webb Space Telescope (JWST) in late 2021 opens a new era 
in the study of galaxy formation and evolution. As an upgrade to its predecessor, 
Hubble Space Telescope (HST), the $\sim 7\times$ light-gathering power, longer 
wavelength coverage, higher sensitivity of infrared imaging and spectroscopy 
equip JWST with unprecedented capability for detecting galaxies at high-redshift. 
Recent analyses based on Cycle 1 early data releases from early release 
observations (ERO) programs, such as ERO SMACS J0723 and Stephans's Quintet, 
as well as on early release science (ERS) programs, such as Cosmic Evolution 
Early Release Science 
\citep[CEERS,][]{finkelsteinCosmicEvolutionEarly2017,finkelsteinLongTimeAgo2022,
finkelsteinCEERSKeyPaper2023} and GLASS James Webb Space Telescope 
Early Release Science \citep[GLASS,][]{treuLookingGLASSJWST2017}, have revealed 
dozens of luminous galaxies at redshift ranging from $z\sim 7$ to $\sim 20$ 
\citep[][]{naiduTwoRemarkablyLuminous2022,
rodighieroJWSTUnveilsHeavily2022,
donnanEvolutionGalaxyUV2022,finkelsteinLongTimeAgo2022,bouwensEvolutionUVLF2023,
finkelsteinCEERSKeyPaper2023,harikaneComprehensiveStudyGalaxies2023,
labbePopulationRedCandidate2023}. 
Theoretical followups have suggested that the high 
UV luminosity and stellar mass of the observed candidates are inconsistent with 
nearly all existing galaxy formation models, including those built with empirical 
methods, semianalytical methods, and hydrodynamic simulations 
\citep{masonBrightestGalaxiesCosmic2023,
naiduTwoRemarkablyLuminous2022,
rodighieroJWSTUnveilsHeavily2022,
harikaneComprehensiveStudyGalaxies2023,
labbePopulationRedCandidate2023,
finkelsteinCEERSKeyPaper2023}. This discrepancy is even in significant tension 
with the upper limit permitted by the current LambdaCDM model
\citep{lovellExtremeValueStatistics2022,
boylan-kolchinStressTestingLambda2023}.
However, given the uncertainties in both galaxy formation models 
and observational measurements, it may be premature to draw any definitive 
conclusions.

Various arguments have been put forward to address these tensions from different 
perspectives. For instance, \cite{masonBrightestGalaxiesCosmic2023} employed 
empirical methods to derive upper limits on the UV luminosity functions at 
redshifts of $z \sim 8 - 20$, and found that the JWST observations are well 
within these limits. A comprehensive study by \citet{harikaneComprehensiveStudyGalaxies2023} 
suggested that the lack of reionization sources at $z \gtrsim 10$ to 
suppress star formation, or a top-heavy initial mass function (IMF) expected 
for Pop-III stars in a lower-metallicity environment exposed to higher cosmic microwave 
background (CMB) temperature, could increase the UV luminosity by a factor of 
$\lesssim 4$, thus marginally resolving the $1\,{\rm dex}$ difference between 
model and observation. \citet{yungAreUltrahighredshiftGalaxies2023} 
arrived at similar conclusions that a modest boost of a factor of $\sim 4$ to the 
UV luminosities, possibly due to a top-heavy IMF, can resolve the $\gtrsim 1\,{\rm dex}$ 
discrepancy between the Santa-Cruz semi-analytical model (SAM) and observations at 
$z \gtrsim 10$. They also suggested another modification to the star 
formation model, assuming a lower stellar feedback strength, to address the tension. 
\citet{kannanMillenniumTNGProjectGalaxy2023} proposed that new processes should 
be incorporated into hydrodynamic simulations to match the observed star formation 
rate density.

Recent observational results by \citet{labbePopulationRedCandidate2023} at 
redshifts of $z \sim 7 - 10$ present even more tensions with current 
hierarchical galaxy formation models under the $\Lambda$CDM cosmology. 
Using double-break selected galaxies from the JWST CEERS sample, they identified 
six candidate galaxies with stellar masses $M_* \geq 10^{10} \msun$, with one 
extreme galaxy having a mass of nearly $10^{11} \msun$. 
To derive redshift and stellar mass, they used the photometry in 10 bands 
from JWST/NIRCam + HST/ACS observations, as well as seven different SED 
fitting code settings (EAZY, Prospector, and five Bagpipes settings). 
Compared to earlier results from HST+Spitzer measurements \citep{stefanonGalaxyStellarMass2021}, 
the cosmic stellar mass density reported by \citet{labbePopulationRedCandidate2023} 
is a factor of $\sim 20$ higher at $z\sim 8$ and a factor of $\sim 1000$ higher 
at $z\sim 9$. If these findings are confirmed by future spectroscopic surveys, 
such as those conducted with JWST/NIRSpec, they present a serious challenge 
to the standard $\Lambda$CDM paradigm, because the observed stellar masses 
exceed the upper limit allowed by the cosmic baryon fraction.

Several theoretical follow-ups have been conducted to explain the observational 
result by \citet{labbePopulationRedCandidate2023}. Using theoretical halo mass 
functions from the Press-Schechter formalism given by \citet{shethLargescaleBiasPeak1999} 
and assuming a maximal star formation efficiency, 
\citet{boylan-kolchinStressTestingLambda2023} showed that the two most massive 
candidates identified by \citet{labbePopulationRedCandidate2023} 
represent a $\sim 3\sigma$ tension. Even taking into account 
uncertainties in stellar mass, sampling and Poisson fluctuation, 
their results still require that the star formation 
efficiency at $z=9$ be $\epsilon_*(z=9) \geqslant 0.57$. 
The tension would be even more severe given that the Sheth-Tormen halo 
mass functions are $20\% - 50\%$ higher than those obtained from N-body 
simulations in the same redshift range. \citet{lovellExtremeValueStatistics2022} 
performed more stringent tests using halo mass functions from N-body simulations 
and Extreme Value Statistics (EVS), and revealed a tension 
at $> 3\sigma$ level between the sample of 
\citet{labbePopulationRedCandidate2023} and the expectation 
from the extreme assumption of $\epsilon_*=1$. Although the results 
may change with updated calibration in the photometry of JWST/NIRCam, 
the qualitative conclusion is expected to remain.

There are several issues with summary statistics obtained from galaxies of extreme
masses. Firstly, uncertainties in the stellar mass estimate can be amplified by 
the steepness of the stellar mass function at the high-mass end, and 
can lead both to larger scatter and bias towards higher values - 
a phenomenon known as Eddington bias 
\citep{eddingtonFormulaCorrectingStatistics1913}. Secondly, 
as errors from different sources may interact non-linearly, incomplete 
modeling of the sources of uncertainties can potentially change the 
statistical result significantly. Finally, the highly skewed and discrete error 
distribution makes it difficult to represent accurately the error distribution 
with centralized and symmetric analytical approximations. 
All these issues have important implications for interpreting 
observational results in terms of theoretical expectations.

In observations, a summary statistic, $s$, is usually represented and plotted 
as $(s_0 - \Delta s_{\rm lower}, s_0 + \Delta s_{\rm upper})$, where $s_0$ is 
interpreted as an estimate of the mean or median of the underlying random variable 
$s$, and $\Delta s_{\rm lower}$ and $\Delta s_{\rm upper}$ represent the standard 
deviation or quantile. However, with biased observations and asymmetric error 
distributions, the $s_0\pm \Delta$ error representation deviates from 
its commonly-assumed physical meaning and may mislead observation-theory 
comparisons. A solution to these issues is the forward approach supported by 
Bayesian theory. \citet{lovellExtremeValueStatistics2022} provided an example by using extreme 
values of stellar mass, instead of direct counting, and forward modeling of 
errors based on halo mass functions and volume sampling to bypass the issues 
described above. However, the volume sampling technique, error modeling, and 
correction for Eddington bias are likely too simplified in their implementation. 
Meanwhile, summary statistics, such as galaxy number density and cosmic stellar 
mass/SFR density, are important scientific products of many HST/Spitzer/JWST 
observations and serve as the entry point for observation-theory comparisons.

In this study, we propose a forward modeling approach of galaxy counting and 
demonstrate the impact of different sources of uncertainties on 
observational results. We begin with an N-body simulation, populate 
halos with galaxies through consecutive transformations, and 
incorporate various sources of uncertainties in these steps. We carefully 
devise both the representation of errors and the method for comparison 
with high-redshift data. Using a constrained simulation that reconstructs 
the assembly history of real clusters of galaxies at $z\sim 0$, 
we provide insights into the low-$z$ descendants 
of the observed high-redshift galaxies.

The paper is organized as follows. In \S\ref{sec:data-and-method}, 
we describe the N-body simulation we use, the method to populate halos with 
galaxies, and the approach to incorporating uncertainties. In 
\S\ref{sec:uncertainties}, we highlight the impact of these uncertainties 
on the interpretation of recent JWST observations at $z \sim 7 - 10$. 
We will show that the tension with the standard $\Lambda$CDM 
can be reduced significantly or even completely eliminated 
by taking these uncertainties properly into account.  
In \S\ref{sec:descendant}, we present the descendant properties of 
high-redshift massive galaxies and discuss their implications for 
observations.


\section{Data and Analysis}
\label{sec:data-and-method}

\subsection{The ELUCID Simulation}
\label{ssec:data-simulation}

\begin{figure*} \centering
    \includegraphics[width=0.86\textwidth]{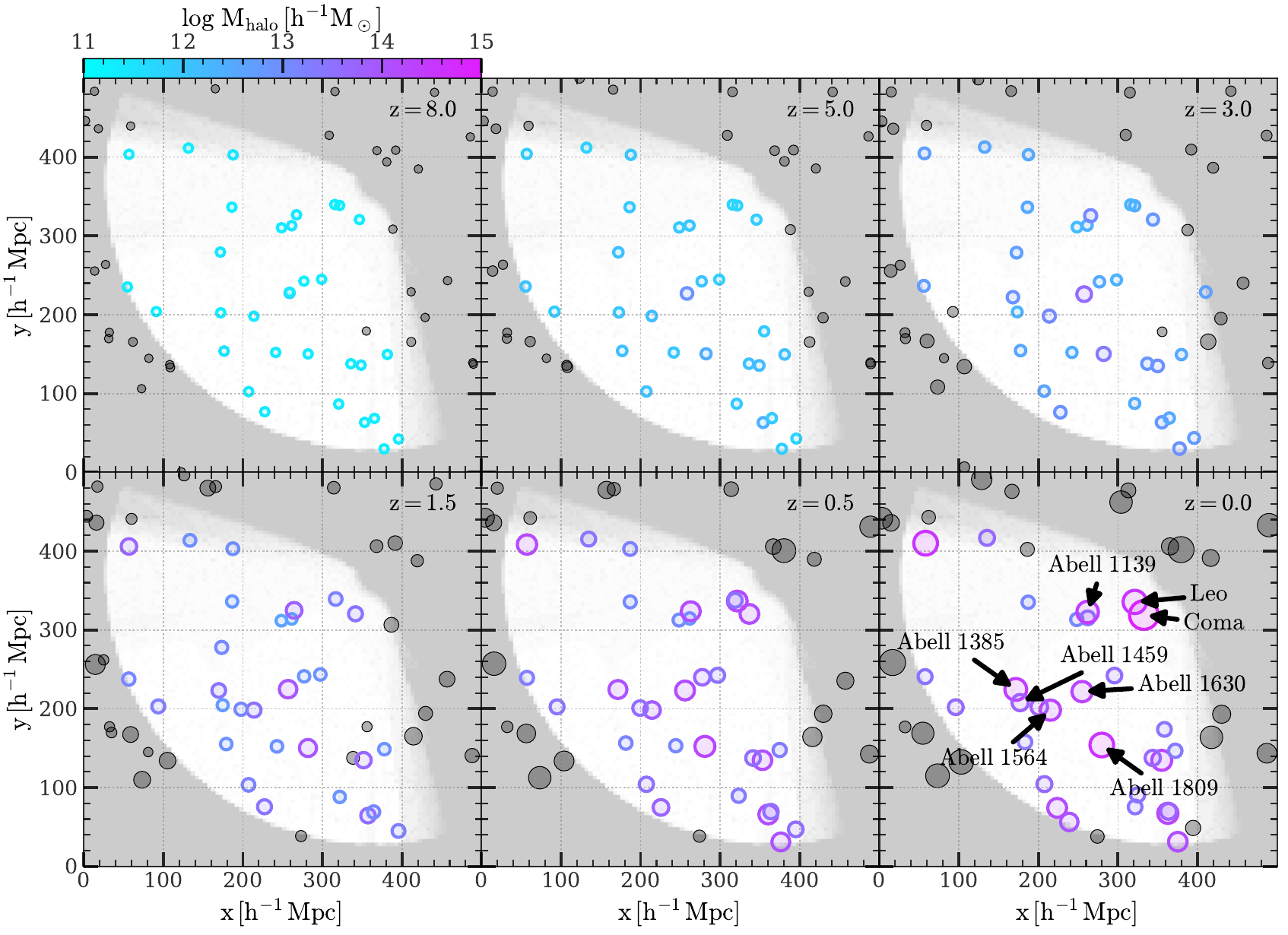}
    \caption{
        Massive halos selected at $z=8$ with $\Mhalo \geqslant 2\times 10^{11}\msun$ 
        from a spatial slice with $50\mpc$ thickness in the constrained simulation 
        of ELUCID (\figbf[top-left panel]) 
        and their descendant halos at lower redshifts (\figbf[other panels]). 
        Descendant halos at $z=0$ in the simulation are matched with observed clusters 
        in the local Universe, as indicated in the \figbf[lower-right panel] with names or Abell indices. 
        The \figbf[gray shaded] area in each panel marks the out-of-reconstruction volume, 
        while the unmasked area is the volume whose density field is constrained by galaxy groups from SDSS. 
        Out-of-reconstruction halos are represented by \figbf[gray circles]. 
        In-reconstruction halos are represented by \figbf[colored circles], 
        whose sizes and colors are scaled according to their masses.  
        For a detailed description of the constrained simulation and halo sample, refer to \S\ref{sec:data-and-method}.
        For a discussion about the evolution of halos, see \S\ref{sec:descendant}.
    }
    \label{fig:spatial_scatter}
\end{figure*}

Throughout this paper, we use ELUCID \citep{wangELUCIDEXPLORINGLOCAL2016}, a 
constrained N-body simulation obtained using the code \softwarenamestyle[L-Gadget], 
a memory-optimized version of \softwarenamestyle[Gadget-2]
\citep{springelCosmologicalSimulationCode2005}. The simulation uses a WMAP5 
\citep{dunkleyFIVEYEARWILKINSONMICROWAVE2009} cosmology with the following parameters: 
$\Omega_{\rm K,0}=0$, $\Omega_{\rm M,0}=0.258$, $\Omega_{\rm B,0}=0.044$, 
$\Omega_{\rm \Lambda, 0}=0.742$, $H_0 = 100\ h\,{\rm km\,s^{-1}\,Mpc^{-1}}$ 
with $h=0.72$, and a spectral index of $n=0.96$ with an amplitude specified by 
$\sigma_8=0.80$ for the Gaussian initial density field. A total of 100 snapshots, 
from redshift $z=18.4$ to $0$, are saved. Halos are identified using the 
friends-of-friends (\softwarenamestyle[FoF]) algorithm 
\citep{davisEvolutionLargescaleStructure1985} with a scaled linking length of $0.2$. 
Subhalos are identified using the \softwarenamestyle[Subfind] algorithm 
\citep{springelPopulatingClusterGalaxies2001, dolagSubstructuresHydrodynamicalCluster2009}, 
and subhalo merger trees are constructed using the \softwarenamestyle[SubLink] 
algorithm \citep{springelCosmologicalSimulationCode2005, 
boylan-kolchinResolvingCosmicStructure2009}. ELUCID has a simulation box with 
a side length of $500\mpc$ and uses a total of $3072^3$ particles to trace the 
cosmic density field. The mass of each dark matter particle is $3.08\times 10^8 \msun$, 
and the mass resolution limit of FoF halos is about $10^{10} \msun$. As we only 
use halos well above this limit to model massive galaxies, 
the relatively low resolution of ELUCID does not affect our conclusions. 
Additionally, because all the uncertainties considered here are 
significant, the slight difference in cosmological parameters 
between WMAP5 and the more recent Planck data does not have a
significant impact on our findings.

The initial condition of ELUCID is reconstructed from real galaxy groups identified 
from the SDSS \citep{yangGalaxyGroupsSDSS2007,yangEvolutionGalaxyDarkMatter2012} by a 
sequence of numerical methods detailed in \citet{wangELUCIDEXPLORINGLOCAL2016}. 
Analysis based on cross-matching the simulated and observed halos at $z \sim 0$ 
showed that more than $95\%$ of massive halos with halo mass 
$M_{\rm halo} \geqslant 10^{14}\msun$ can be recovered in the constrained 
simulation with a $0.5\,{\rm dex}$ tolerance on the error of halo mass and a 
$4\mpc$ tolerance on the error of spatial location. These features of ELUCID 
enable not only the statistical study of halo and galaxy populations within it, 
but also a halo-by-halo comparison to massive clusters in the real universe. 
In \S\ref{sec:uncertainties}, we use ELUCID to statistically infer uncertainties 
in measuring the abundance of extremely massive galaxies at $z = 7 \sim 10$, 
and in \S\ref{sec:descendant}, we use ELUCID to study the assembly history of 
several real massive clusters in the local universe. The bottom-right panel of 
Fig.~\ref{fig:spatial_scatter} shows some examples of the matched clusters 
in a slice of the simulation volume, including two massive clusters,
Coma and Leo, and a number of Abell clusters 
\citep{abellCatalogRichClusters1989,strubleCompilationRedshiftsVelocity1999}.

\subsection{Method to Evaluate Uncertainties in Galaxy Counting}
\label{ssec:method-uncertainties}

Here we take ELUCID as the input, populate dark matter halos with galaxies 
and estimate the uncertainties in counting galaxies for surveys at high $z$.
Our goal is to estimate the uncertainty of a given summary statistic, 
$s_*$, of galaxies in a sub-volume whose geometry is consistent with the survey 
in question, starting from the sample $S_{\rm halo}$ of all halos at a given 
redshift $z$ in ELUCID. To achieve this, we decompose the transformation from 
$S_{\rm halo}$ to $s_*$ into a chain of four steps based on our understanding 
of galaxy formation in the $\Lambda$CDM paradigm:
\begin{equation}
    s_* = \mathcal{T}_{\rm stat} \circ
        \mathcal{T}_{M_*} \circ
        \mathcal{T}_{M_{\rm halo}} \circ 
        \mathcal{T}_{\rm CV} (S_{\rm halo})\,,
\end{equation}
where $\mathcal{T}_{\rm CV}$ denotes the volume-based sub-sampling of halos, 
$\mathcal{T}_{M_{\rm halo}}$ denotes the determination of the mass 
$M_{\rm halo}$ for each halo in the sub-sample, $\mathcal{T}_{M_*}$ denotes the 
halo-to-galaxy mapping, and $\mathcal{T}_{\rm stat}$ denotes the statistical 
function that extracts the summary statistic from the galaxy sample. 
We will describe numerical implementations of these steps in detail later.

As discussed in \S\ref{sec:intro}, there are multiple uncertainties that are 
critical in estimating $s_*$. Our decomposition strategy described above 
ensures that each type of uncertainty can be physically modeled in the
relevant step, and that the forward incorporation of all steps naturally 
propagates all uncertainties into the total uncertainty of $s_*$. In addition, 
this also allows us to quantify contributions of individual uncertainties 
by incrementally adding them into the chain.

In principle, the counts of halos are noisy because massive 
objects are rare in a small volume by definition. This sampling uncertainty 
is further enhanced by other sources of uncertainty in the halo-to-galaxy 
mapping due to the increased steepness of the Schechter function 
toward the high-mass end. Additionally, as the sample 
size goes below $\sim 100$, the discreteness in galaxy counts and its skewness 
must be carefully taken into account \citep[e.g.,][]{trentiCosmicVarianceIts2008}. 
To address these issues, we deliberately choose the cumulative cosmic stellar mass 
density, $s_* = \rho_*(>M_*)$, as the summary statistic when comparing the theoretical 
prediction with observational data \citep[see also][for the same usage]{labbePopulationRedCandidate2023,
lovellExtremeValueStatistics2022,boylan-kolchinStressTestingLambda2023}. 
We describe the uncertainty of $s_*$ using the probability distribution, $P(s_*)$, 
instead of summary statistics with compressed information, such as the average, 
variance, and quantile. If desired, these compressed quantities can be estimated 
from $P(s_*)$.

In the following, we specify each of the transformations involved in the mapping 
from halos to galaxy statistic and describe uncertainties injected into them.

{\it Volume sampling operator}: 
The transformation $\mathcal{T}_{\rm CV}$, implemented by drawing a sub-volume 
from the periodic box of ELUCID and filtering out all halos from $S_{\rm halo}$ 
outside the selected volume, constitutes the volume sampling operator. 
The uncertainty in the sampling is naturally incorporated due to the 
fluctuation of the density field seeded by the initial condition. Throughout 
this paper, we refer to the uncertainty introduced by this sampling step 
collectively as the cosmic variance (CV). We do not subtract the Poisson 
shot noise from this uncertainty because its exact effect on $\rho_*(>M_*)$ is 
not analytically traceable, and any inaccuracy in the approximation 
can change the extreme statistics significantly. When estimating CV for a real 
survey, the sub-volume must be chosen to have a consistent shape with the 
survey, especially for a survey with a pencil-beam-like space coverage. This is 
because a beam-shaped volume passes many different environments and is thus 
less likely to be affected by a single over- or under-density region 
in comparison to a cube-shaped volume, as discussed by 
\citet{trentiCosmicVarianceIts2008} and \citet{mosterCOSMICVARIANCECOOKBOOK2011}.

{\it Halo mass estimator}: 
The operator $\mathcal{T}_{\rm M_{\rm halo}}$ obtains halo masses from the 
sample of halos obtained from the previous step. In this paper, we define the 
halo mass, $M_{\rm halo}$, as the total mass of all the dark 
matter particles bound to the central subhalo when  
mapping the halo to its central galaxy. We have verified that the 
definition of halo mass has a negligible effect on our results for 
halos with $M_{\rm halo} \geqslant 10^{10.5} \msun$ in comparison 
to other more significant uncertainties. In the $\Lambda$CDM paradigm, the 
combination of halo mass and cosmic baryon fraction, 
$f_{\rm b}=\Omega_{B,0}/\Omega_{M,0}$, sets a natural upper limit 
on the baryon mass that can be converted into stars. However, backsplash halos
\citep[e.g.,][]{baloghOriginStarFormation2000,wangDistributionEjectedSubhalos2009,
wetzelGalaxyEvolutionGroups2014,diemerFlybysOrbitsSplashback2021,
wangDissectTwohaloGalactic2023}, 
whose dark matter has left the host halo while its gas   
may remain in the host halo, can increase the baryon-to-dark matter 
ratio in the host halo, thereby lifting the upper limit. 
To model this uncertainty, we add the mass of backsplash halos to the  
$M_{\rm halo}$ of the host halo, and we use this new mass 
as an ``effective halo mass'' of the host to estimate the upper limit 
on the available baryons. Our tests show that the added mass is, on average, 
about $50\%$ of the original mass for the most massive halos at 
$z=7 \sim 10$, and less significant for less massive halos.

{\it Stellar mass estimator}: 
The transformation $\mathcal{T}_{\rm M_*}$ assigns each halo a stellar mass, 
$M_*$, for its central galaxy, based on the halo mass obtained from the 
previous step. Assuming a constant star formation efficiency 
$\epsilon_*$, we model this mapping as:
\begin{equation}
    M_*=\epsilon_* f_{\rm b} M_{\rm halo}.
\label{eq:m_star_estimator}
\end{equation}
The simplification using a constant star formation efficiency is motivated 
by observational measurements from a HST+Spitzer sample   
by \citet{stefanonGalaxyStellarMass2021}, where the star 
formation efficiency is found to be flat at the high-halo-mass end 
and shows little evolution in $z \sim 6 - 10$. Recent empirical models 
\citep[e.g.,][]{behrooziUniverseMachineCorrelationGalaxy2019}, semi-analytical models 
\citep[e.g.,][]{yungAreUltrahighredshiftGalaxies2023}, and hydrodynamic 
simulations \citep[e.g.,][]{kannanMillenniumTNGProjectGalaxy2023} do not 
seem capable of producing such high-mass galaxies at high $z$, 
suggesting a high value of $\epsilon_*$ must be assumed to match 
the observational data.

However, in real observations, the measurement of $M_*$ suffers from systematic 
and random errors. Each assumption on the IMF, star formation history, 
dust attenuation, and photometric redshift can introduce significant amounts  
of uncertainty into the SED fitting
\citep[see][for a comprehensive review of techniques and uncertainties]{
conroyModelingPanchromaticSpectral2013,
behrooziCOMPREHENSIVEANALYSISUNCERTAINTIES2010,
stanwayApplicationsStellarPopulation2020}. 
Indeed, \citet{harikaneComprehensiveStudyGalaxies2023} and 
\citet{finkelsteinCEERSKeyPaper2023} found that these assumptions 
can introduce significant uncertainty in the results of JWST galaxies. 
\citet{vanmierloNoNeedExtreme2023} performed a test on a $z\sim 7$ massive galaxy 
with different SED fitting codes, and found a $0.76\,{\rm dex}$ 
systematic offset in stellar mass among the five codes adopted. Similarly, 
\citet{labbePopulationRedCandidate2023} performed SED fittings with seven 
different methods, and found $\sim 0.1 - 0.3\,{\rm dex}$ random 
error in stellar mass by using a given method, and 
$\sim 0.2 - 1\,{\rm dex}$ systematic difference between different methods. 
For some galaxies, the systematic difference was found to be as 
large as $2\,{\rm dex}$. To incorporate such uncertainties, we introduce two free 
parameters in the mapping from $M_{\rm halo}$ to $M_*$: a constant bias factor 
$b_{M_*}$ and a normally distributed random error with zero mean and a standard 
deviation of $\sigma_{M_*}$. A typical value for these parameters is about 
$0.1$ to $0.5$ dex in low-redshift measurements, depending on galaxy sample in use
and the model choices made \citep[e.g.,][]{liDistributionStellarMass2009,
conroyModelingPanchromaticSpectral2013}, 
and they are expected to be at least as large for the high-$z$ galaxies 
concerned here.

{\it Statistical operator}: 
The final transformation, $\mathcal{T}_{\rm stat}$, is a statistical function 
that compresses the sample of $M_*$ into $s_*=\rho_*(>M_*)$. This is simply 
a cumulative histogram weighted by $M_*$. No error should be added in this step.

By applying all four steps, we obtain an estimate for the stellar mass density 
that mimics real observations. To obtain a sample of $\rho_*$, we repeatedly 
select sub-samples from the entire volume of ELUCID and obtain $\rho_*$
from each of them. From this sample of $\rho_*$, we numerically compute the 
probability distribution $P(\rho_*)$ and other summary 
statistics based on $\rho_*$. Finally, we make a model-observation comparison 
using $P(\rho_*)$ to obtain the probability of observing a particular value of $\rho_*$.

In some of our analyses, we use halo and galaxy properties 
other than those defined above. When linking halos across redshifts, we are 
more concerned about properties of a halo as a whole, rather than 
stellar contents within it. In this case, we adopt the ``tophat'' halo mass, 
calculated using dark matter particles enclosed in a virial 
radius within which the mean density is equal to that of the spherical 
collapse model \citep{bryanStatisticalPropertiesXRay1998}. When demonstrating 
the absolute number of halos/galaxies in bins of given masses, we use the 
un-weighted, un-normalized histogram and its cumulative version, instead of the 
distribution density. We will clarify their usage in the description of our results.

\begin{figure*} \centering
    \includegraphics[width=0.73\textwidth]{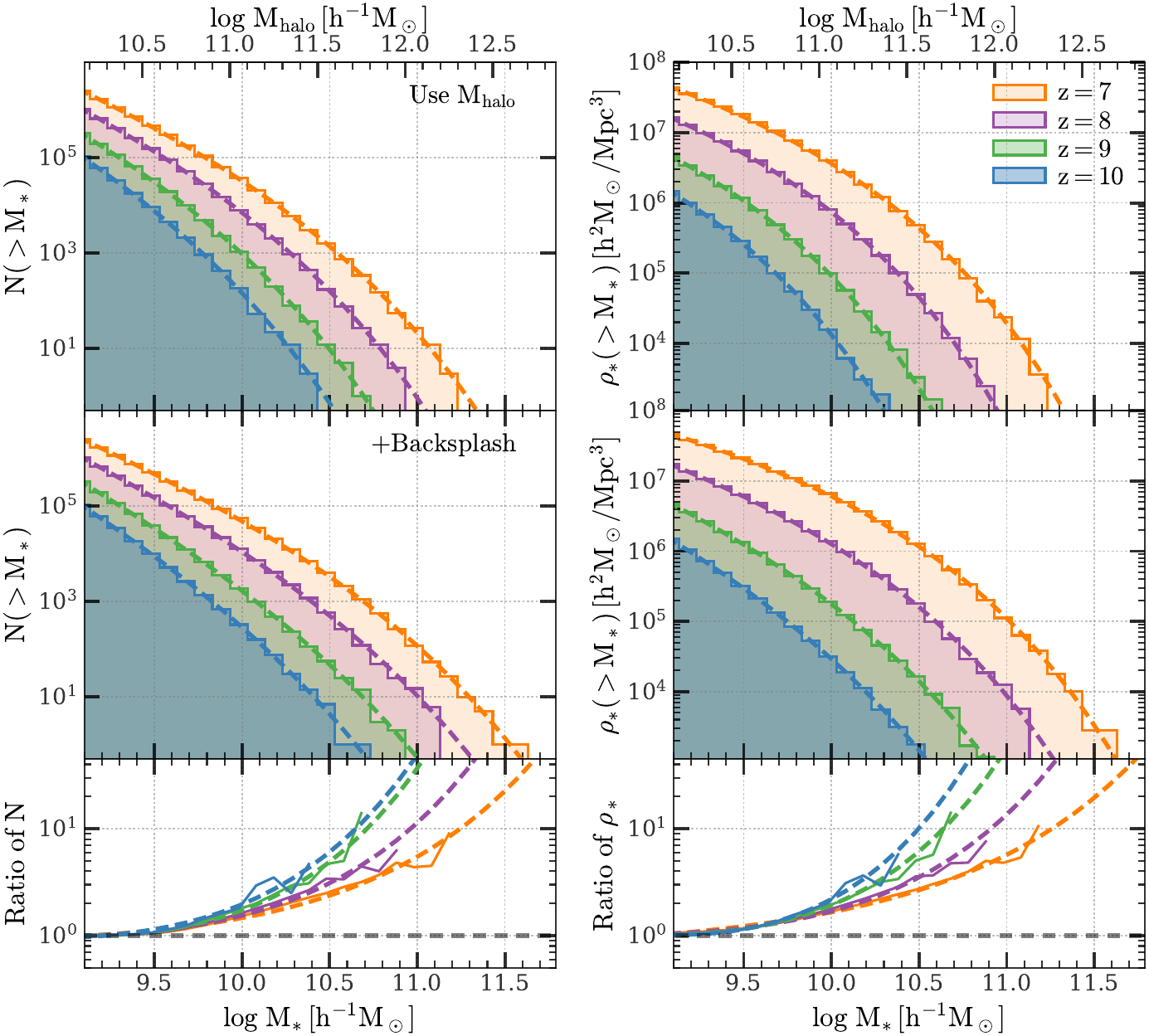}
    \caption{
        Cumulative distributions of halo mass $\Mhalo$ of halos and stellar mass $M_*$ of central galaxies. 
        The \figbf[left column] shows the cumulative number in the entire simulation volume of ELUCID, 
        while the \figbf[right column] shows the cumulative stellar mass density. 
        In the \figbf[first row], the bound dark matter mass of the central subhalo is used, 
        while in the \figbf[second row], the mass brought in by backsplash halos is added. 
        The \figbf[third row] shows the ratio of the number/mass density with added backsplash mass to that without backsplash mass. 
        In each panel, \figbf[solid lines] are obtained from simulated halos, 
        while \figbf[dashed lines] are obtained by fitting with Schechter functions. 
        Results at $z = 7$, $8$, $9$ and $10$ are shown by \figbf[different colors] as indicated in the \figbf[top-right corner]. 
        A star formation efficiency $\epsilon_*=0.5$ is assumed in the conversion from halo mass to stellar mass. 
        The significant enhancement in the number and mass densities at the high-stellar-mass end has non-negligible 
        implications in the interpretation of the observed summary statistics. 
        For further details, refer to \S\ref{ssec:uncertainties-expected}.
    }
    \label{fig:backsplash}
\end{figure*}

The top-left panel of Fig.~\ref{fig:spatial_scatter} shows a sample of halos 
with $M_{\rm halo} \geqslant 2\times 10^{11} \msun$ at $z=8$ in a slice of the 
simulation box. At this redshift, there are already a large number of massive halos. 
Assuming a large but reasonable star formation efficiency, $\epsilon \sim 0.5$, 
these halos are capable of hosting massive galaxies with 
$M_* \gtrsim 10^{10} \msun$ that have been identified by JWST 
\citep[e.g.,][]{rodighieroJWSTUnveilsHeavily2022,labbePopulationRedCandidate2023}. 
The two panels in the first row of Fig.~\ref{fig:backsplash} show the cumulative 
number density, $N(>M_*)$, and the cumulative cosmic stellar mass density, $\rho_*(>M_*)$, 
respectively, as functions of $M_*$ for galaxies in the entire simulation volume of 
ELUCID at snapshots from $z=7$ to $10$, with $M_*$ obtained by assuming 
a constant star formation efficiency $\epsilon_* = 0.5$. 
More than 1000 galaxies with $M_* \geqslant 10^{10} \msun$ 
can be found in the entire volume of ELUCID. In the following section, we examine  
whether or not the number of massive objects observed by JWST can be accommodated 
by the simulation.

\section{Quantifying Uncertainties in Galaxy Counting}
\label{sec:uncertainties}
\subsection{Expected Uncertainties from Different Sources}
\label{ssec:uncertainties-expected}

\begin{figure*} \centering
    \includegraphics[width=0.62\textwidth]{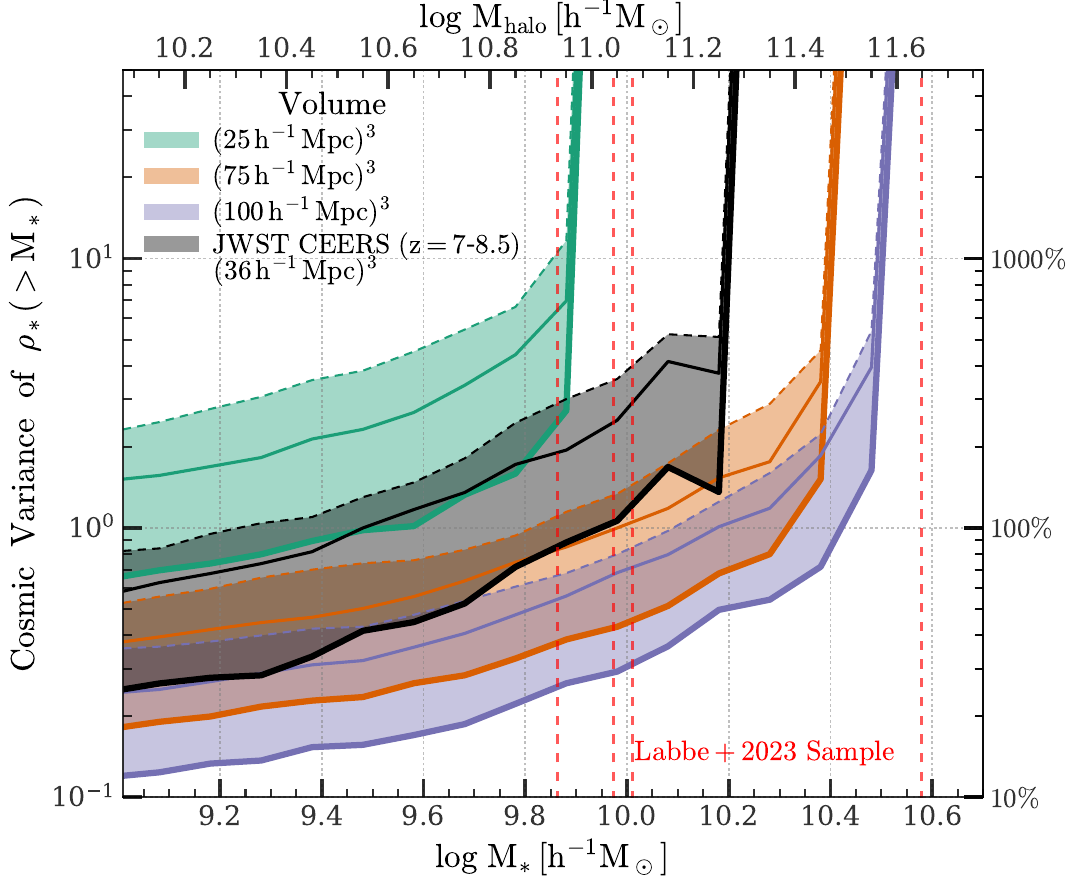}
    \caption{
        Cosmic variance on cumulative cosmic stellar mass density, $\rho_*(>M_*)$, at $z=8$ 
        estimated from the ELUCID simulation. \figbf[Colored lines] show the cosmic variance 
        in boxes with given volumes indicated in the top-left corner, 
        while \figbf[black lines] show the cosmic variance in a volume whose size
        and shape are consistent with the JWST CEERS program at $z = 7$-$8.5$. 
        For each case, \figbf[thick solid], \figbf[thin solid] and \figbf[dashed] lines 
        represent the $1$, $2$, and $3\,\sigma$ uncertainties, respectively, due to cosmic variance, 
        which are obtained by randomly drawing 1000 subvolumes from the ELUCID simulation. 
        The fractional value is labeled on the \figbf[left axis], while the percentage value is 
        labeled on the \figbf[right axis]. 
        \figbf[Vertical red dashed lines] indicate the masses at which 
        $\rho_*(>M_*)$ is measured by \citet{labbePopulationRedCandidate2023} 
        using the JWST CEERS sample consisting of four massive galaxies. 
        A star formation efficiency $\epsilon_*=0.5$ is assumed in the conversion 
        from halo mass to stellar mass. These results suggest significant mass 
        dependence of cosmic variance and its capability to boost the spatial density 
        of high-mass-end galaxies by more than an order of magnitude. 
        For further details, refer to \S\ref{ssec:uncertainties-expected}.
    }
    \label{fig:cv_z8}
\end{figure*}

\begin{figure*} \centering
    \includegraphics[width=0.73\textwidth]{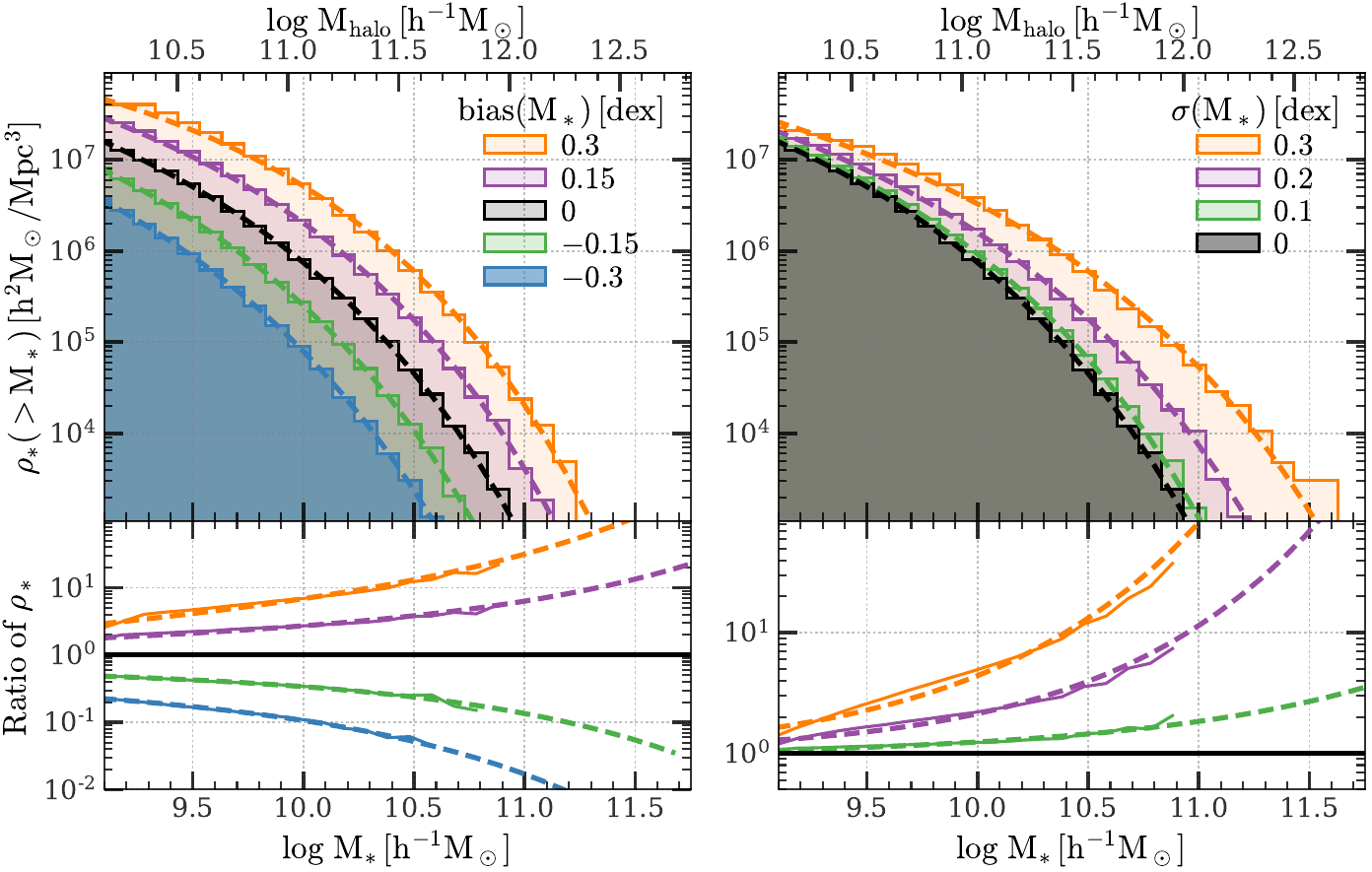}
    \caption{
        The impact of systematic and random errors in measuring stellar mass on 
        cosmic stellar mass density. The \figbf[left column] represents different 
        systematic biases assumed and indicated in the upper right corner, 
        while the \figbf[right column] represents different random errors assumed 
        and indicated in the upper right corner. 
        The \figbf[first row] displays the cumulative cosmic stellar mass density, 
        while the \figbf[second row] shows the ratio of the density with systematic/random 
        error to that without error. 
        \figbf[Solid lines] are obtained from simulated halos, 
        while \figbf[dashed lines] are obtained by fitting with Schechter functions. 
        All results are derived from ELUCID at $z=8$. 
        A star formation efficiency $\epsilon_*=0.5$ is assumed in the conversion from halo mass to stellar mass. 
        These findings highlight the importance of accurately estimating stellar mass 
        to measure the abundance of massive galaxies. 
        For further details, refer to \S\ref{ssec:uncertainties-expected}.
    }
    \label{fig:m_star_systematics}
\end{figure*}

Using the halo-to-galaxy transformations defined in the previous section, 
we now quantify the effect of their uncertainties on the estimate of the stellar 
mass density, $\rho_*(>M_*)$. We first describe the effect of each 
uncertainty separately by zeroing out the other uncertainties. We then  
combine them to see their synergistic effect.

The second and third rows of Fig.~\ref{fig:backsplash} show the effect of mass 
enhancements introduced by backsplash halos on the transformation 
$\mathcal{T}_{M_{\rm halo}}$, where the ``effective halo mass'' is used 
to predict the stellar mass in it. 
To compare the results at the massive end, where the number/mass 
density drops down to the limit of the simulation, we fit the 
histogram with a  Schechter function and extrapolate it to higher mass. 
Compared with the results without backsplash, the $N(>M_*)$ or $\rho_*(>M_*)$ 
for galaxies with $M \lesssim 10^9 \msun$ do not change. However, 
the effect of backsplash is more significant for galaxies of higher mass.
The density can reach $\sim 200\%$ of its original value at 
$M_* \sim 10^{10} \msun$ for all redshifts shown, and go beyond 
$1000\%$ at the high-mass end. This mass-dependency of the 
backsplash enhancement is the consequence of two effects. The first 
is that more massive halos are, on average, embedded in denser environments 
where high-speed close encounters are more frequent and 
the fraction of mass brought in by backsplash halos is higher. 
Second, the Schechter function for halo mass distribution 
declines exponentially at the high-mass end, so that 
the change in the halo mass is magnified in the halo mass function. 
It is not yet clear whether or not the baryon component of a 
backsplash halo can be effectively acquired by the host halo 
after the dark matter component is ejected, and whether or not the 
acquired gas can form stars effectively. Future detailed hydrodynamic 
simulations are needed to clarify the ambiguity.

Fig.~\ref{fig:cv_z8} shows the effect of cosmic variance (CV) on $\rho_*(>M_*)$ 
introduced by the sub-sampling operator $\mathcal{T}_{\rm CV}$:
\begin{equation}
    CV(\rho_* | p) = \frac{\rho_{*, 0.5(1+p)}}{\rho_{*,0.5}} - 1.
\end{equation}
Here, $p \in [0,1]$ is the target fraction of the probability mass centralized at the 
median, $\rho_{*, 0.5(1+p)}$ is the quantile at the percentile $0.5(1+p)$, and 
$\rho_{*,0.5}$ is the median. We deliberately avoid using the average 
and standard deviation in the measurement of error, as they are not stable nor 
informative for a highly skewed distribution.

Throughout this paper, we use 1, 2 and 3$\sigma$ to denote cases where 
$p=0.68$, $0.95$ and $0.99$ are used for the quantiles, respectively. 
The distribution of $\rho_*$ is obtained numerically by sub-sampling the 
periodic box of ELUCID with a given volume, and the quantiles are estimated 
from the distribution.
Here, we take the $z=8$ snapshot as an example and show CV for different 
volumes and for different thresholds of $M_*$. Solid, dashed, 
and dotted colored lines in Fig.~\ref{fig:cv_z8} show 1, 2, and 3$\sigma$ CVs 
for a cubic volume of a given size. For low-mass galaxies with 
$M_* \sim 10^{9} \msun$, the $1\sigma$ CV is estimated to be about 
$10\%$ for a volume of $(100\mpc)^3$ and increases monotonically 
to $\sim 70\%$ for a volume of $(25\mpc)^3$. 
This is consistent with the results obtained from the cosmic variance estimator 
given by \citet{chenELUCIDVICosmic2019} for low-$z$ samples. 
Black lines show the CV expected for JWST CEERS at $z=7 - 8.5$, assuming 
a $38\,{\rm arcmin}^2$ effective sky area as used by \citep{labbePopulationRedCandidate2023}. 
The sub-sampling in ELUCID is conducted using beam-shaped volumes with a size 
of $(36\mpc)^3$ and a tangential-to-normal aspect ratio of $12 / 342$.
The 1, 2, and 3$\sigma$ CVs are estimated to be $25\%$, $60\%$, and $80\%$ 
for low-mass galaxies. The moderate effect of CV on this mass scale indicates 
that the density of $M_* \lesssim 10^9\msun$ galaxies can be estimated reliably 
in JWST CEERS if other uncertainties are well controlled.

However, the effect of CV has a significant dependence on stellar mass, 
with uncertainty reaching $100\%$ at higher stellar masses and eventually becoming 
divergent at the highest-mass end, regardless of the survey volume. 
The point of divergence shifts rightward as the volume increases, 
as it emerges when the median of $\rho_*(>M_*)$ approaches zero. 
In Fig.~\ref{fig:m_star_systematics}, the vertical lines represent the 
masses at which $\rho_*(>M_*)$ is measured by 
\citet{labbePopulationRedCandidate2023} using the four
massive galaxies from JWST CEERS. Three of these lines intersect with the 
black solid line at CV $\sim 100\%$, suggesting that the CV is 
a significant effect for this small sample. The rightmost line, obtained 
from a galaxy with the most extreme stellar mass of $10^{10.89}\Msun$ at 
$z \sim 7.48$, is located in the divergent region of CV, indicating an 
incredible effect of uncertainty in the statistics drawn from this single galaxy. 
Since the divergence of CV is tightly related to the vanishing of the 
median $\rho_*(>M_*)$ by its definition, it is more informative to model 
$\rho_*(>M_*)$ forwardly, by directly predicting the distribution 
$P[\rho_*(>M_*)]$ and the probability of observing a value 
of $\rho_*(>M_*)$. We will demonstrate this later in this section.

Figure ~\ref{fig:m_star_systematics} illustrates the impact of the uncertainty 
in the stellar mass estimator, $\mathcal{T}_{M_*}$. To demonstrate the effect, 
we take all halos at the $z=8$ snapshot of ELUCID as an example. In the left panel, 
we show the change of $\rho_*(M_*)$ when different levels of systematic error
i.e. bias, are added to the stellar mass predicted by Eq.~\ref{eq:m_star_estimator}. 
A negative (positive) bias naturally decreases (increases) $\rho_*(M_*)$, as it 
is equivalent to a lower-left (upper-right) shift of the distribution function. 
However, when considering the ratio of the biased distribution to the unbiased one, 
a stellar-mass dependency is observed. This is similar to the mass dependency of CV, 
where a small perturbation to the transformation is magnified significantly owing 
to the steepness of the Schechter function at the high-stellar-mass end. 
The overall outcome is an increased effect of the uncertainty in the stellar 
mass estimate at the high-mass end, and a divergence occurs when the unbiased 
galaxy number $N(>M_*)$ approaches zero.

In the right panel of Fig.~\ref{fig:m_star_systematics}, we show the effect 
of random error, known as the Eddington bias \citep{eddingtonFormulaCorrectingStatistics1913}, 
in the stellar mass estimate. Here, we model the error as a Gaussian random 
variate with zero mean and varying standard deviation, $\sigma_{M_*}$. 
Lines with different colors represent results with different $\sigma_{M_*}$, 
and are compared to the result assuming no error. Once again, to enable a 
comparison in the full mass range, we fit the histogram 
to a Schechter function and extrapolate it to the uncovered 
mass range. As the distribution of $M_*$ has a negative derivative, 
a symmetric noise in $M_*$ produces an 
asymmetric effect on the number counts of galaxies. As the 
derivative becomes more negative toward the high-stellar-mass end, the 
effect of this type of noise becomes more significant. We do see this
expected behavior in the panel, where the histogram is lifted everywhere 
by a nonzero error of $M_*$. With a moderate 
random error of 0.2 dex (0.3 dex), the increase of $\rho_*(M_*)$ is 
$\sim 100\%$ at $M_* \sim 10^{9.2}\msun$ ($10^{9.7}\msun$), and becomes 
divergent when the number of galaxies $N_{>M_*}$ approaches zero.

As suggested by \citet{lovellExtremeValueStatistics2022} and 
\citet{boylan-kolchinStressTestingLambda2023}, an assumption of $\epsilon_* \sim 1$ 
still results in a $\sim 3\sigma$ tension in $\rho_*(>M_*)$ with recent JWST 
observations, especially when the massive galaxies at $z=7-10$ 
from the JWST CEERS sample \citep{labbePopulationRedCandidate2023} are included. 
With all the uncertainties introduced above, and the fact that all of them 
have a strong effect on the statistics drawn from samples of small size, 
it is likely that the observational results can be reproduced by the theory 
with much less tension, as demonstrated in the following.

\subsection{Implications for JWST Observations}
\label{ssec:uncertainty-in-jwst}

\begin{figure*} \centering
    \includegraphics[width=0.89\textwidth]{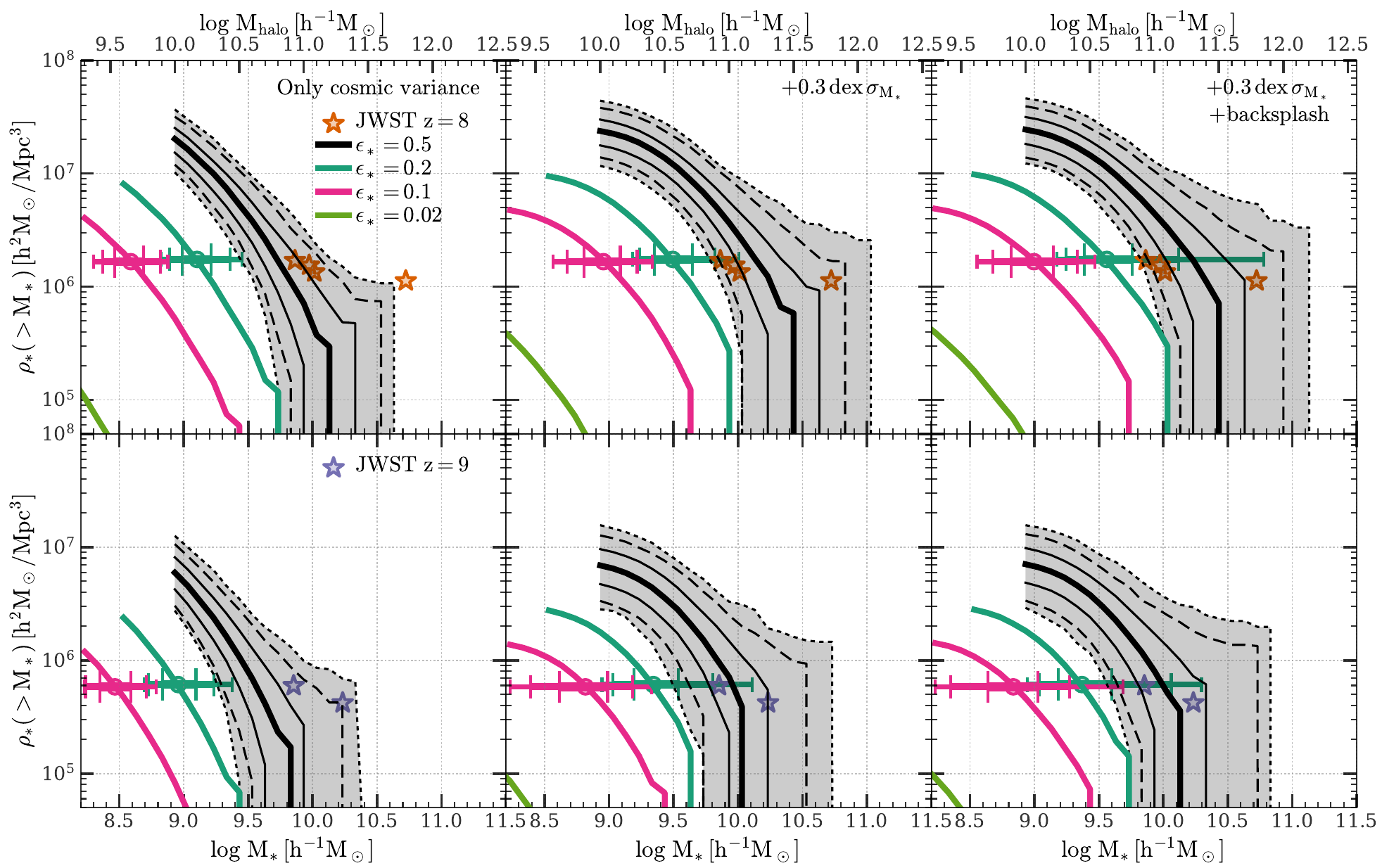}
    \caption{
        A comparison of the cumulative cosmic stellar mass density at $z=8$ (\figbf[first row]) 
        and $z=9$ (\figbf[second row]) with the results obtained from the JWST CEERS 
        sample \citep{labbePopulationRedCandidate2023}. The different uncertainties, 
        including cosmic variance, random error in measuring stellar mass, and the 
        mass enhancement by backsplash systems, are incrementally added and shown 
        \figbf[from left to right columns]. 
        In each panel, the \figbf[thick solid] line represents the median, 
        while the {\bf thin solid, dashed and dotted} lines represent the 
        1, 2, and 3 $\sigma$ ranges, respectively, of the stellar mass density. 
        The volume used in the estimate of cosmic variance is chosen to be consistent 
        with the CEERS sample at the corresponding redshift. For black lines, a star 
        formation efficiency of $\epsilon_*=0.5$ is assumed in the conversion from 
        halo mass to stellar mass (refer to \S\ref{sec:data-and-method} for details). 
        For references, the \textbf{colored curves} show the results obtained by 
        assuming $\epsilon_*=0.2$, $0.1$, and $0.02$, respectively. The horizontal 
        error bars indicate the 1, 2, and 3 $\sigma$ ranges at the stellar mass 
        density corresponding to the leftmost observational data point in the same panel.
        These results suggest that the 
        observed high-mass sample in the recent JWST survey can be safely covered 
        by $2\sigma$ with a reasonable star formation efficiency of $\epsilon_*=0.5$. 
        The inclusion of backsplash halos may further reduce the tension to 
        $\sim 1 \sigma$. 
        For further details, refer to \S\ref{ssec:uncertainty-in-jwst}.
    }
    \label{fig:mass_func_for_jwst_obs}
\end{figure*}

Fig.~\ref{fig:mass_func_for_jwst_obs} shows the cumulative cosmic stellar mass 
density $\rho_*(M_*)$ as a function of the limiting stellar mass $M_*$ for two 
different redshifts, $z \sim 8$ and $z \sim 9$, where six extremely massive 
galaxy candidates are identified and $\rho_*(>M_*)$ estimated  
by \citet{labbePopulationRedCandidate2023}. We demonstrate effects of 
different sources of uncertainties by incrementally adding them into 
the transformations, 
$\mathcal{T}_{\rm CV}$, $\mathcal{T}_{\rm M_{halo}}$, and $\mathcal{T}_{\rm M_*}$, 
respectively, in the mapping from halo sample, $S_{\rm halo}$, to the galaxy statistic, 
$\rho_*(>M_*)$, as introduced in \S\ref{ssec:method-uncertainties}. 
The JWST detections are overplotted for comparison.

Black lines with gray shading in the left column of Fig.~\ref{fig:mass_func_for_jwst_obs} 
show the median and different $\sigma$ ranges derived from the probability distribution 
$P[\rho_*(>M_*)|M_*\,,\epsilon_*=0.5\,,{\rm CV}]$ 
when cosmic variance is incorporated by volume sub-sampling and a star formation 
efficiency $\epsilon_*=0.5$ is adopted to convert halo mass to stellar mass. 
The results shown here are consistent with those obtained by \citet{lovellExtremeValueStatistics2022} 
and \citet{boylan-kolchinStressTestingLambda2023} that some data points are 
outside the $\sim 3\sigma$ range. The exact tension level is slightly 
different, likely because of their simplification in the error calculation with 
analytical approximations, the difference in the cosmologies adopted, and 
the difference in the version of galaxy data. The point obtained by us from a 
$z \sim 8$ extreme is the only one that lies outside the $3\sigma$ range, 
while the $z \sim 9$ extreme marginally touches the $2\sigma$ boundary. 
Other points are all contained within the $2\sigma$ boundary.

As demonstrated in \S\ref{ssec:uncertainties-expected} and \S\ref{fig:m_star_systematics}, 
Eddington bias is able  to lift the galaxy number by more than an order of magnitude, 
which may help explain the outliers we see here. The middle column of 
Fig.~\ref{fig:mass_func_for_jwst_obs} shows the probability distribution 
$P[\rho_*(>M_*)|M_*,\,\epsilon_*=0.5,\,{\rm CV},\,{\rm \sigma_{M_*}}=0.3\,{\rm dex}]$ 
when a Gaussian random error with a conservative 0.3 dex standard deviation is 
added to the estimate of $\log\,M_*$. A significant shift of the median 
$\rho_*(>M_*)$ and a significant broadening of the quantile ranges can be seen after 
the incorporation of this uncertainty. The $z=8$ extreme is now safely contained 
within the 2$\sigma$ range of $\rho_*(>M_*)$, and all the $z=9$ points are 
contained by the $1\sigma$ range. Thus, once the cosmic variance and 
the Eddington bias are included, there is no significant tension 
between the $\Lambda$CDM paradigm and the JWST observations, if the star 
formation efficiency can reach $0.5$ at these redshifts.

The right column of Fig.~\ref{fig:mass_func_for_jwst_obs} shows the results 
for $\rho_*(M_*)$ when the third source of uncertainty, the backsplashed mass, 
is taken into account and added into the halo mass estimator, $\mathcal{T}_{M_{\rm halo}}$. 
Unlike cosmic variance and random error in the stellar mass estimate, this effect 
always increases halo mass and thus provides a larger effective baryon mass 
for star formation. The outcome is obvious: all the data points at 
$z \sim 8$ and $9$ are now below the upper $2\sigma$ quantile
once a star formation efficiency $\epsilon_*=0.5$ is assumed.
At $z \sim 8$, three of the four data points actually lie close to the 
lower $2\sigma$ quantile, while the most massive is close 
to the upper $1\sigma$ quantile.

The colored curves in Fig.~\ref{fig:mass_func_for_jwst_obs} show the results 
obtained by considering lower star formation efficiencies, 
$\epsilon_*=0.2$, $0.1$, and $0.02$, which are more realistic expected from low-z 
observations.
Each curve is overplotted with a series of horizontal error bars, which indicate 
the $1\sigma$, $2\sigma$, and $3\sigma$ ranges at the stellar mass density 
corresponding to the leftmost observational data point in the same panel. 
When solely considering cosmic variance and assuming $\epsilon_* = 0.2$, none of 
the observational data points fall within the $3\sigma$ ranges. However, when 
introducing a standard deviation of 0.3 dex to the estimate of $\log,M_*$, the 
leftmost data points for both the $z=8$ and $z=9$ observations fall within the 
$3\sigma$ range. Furthermore, when incorporating the backsplash effect, 
these observations shift into the $2\sigma$ ranges. These findings suggest that 
a star formation efficiency of $\epsilon_*=0.2$ can marginally account for the 
high stellar mass densities obtained from the CEERS sample. Nevertheless, when 
considering more common values for the low-z Universe, for example, 
$\epsilon_*=0.1$ and $\epsilon_*=0.02$ \citep[e.g.,][]{yangConstrainingGalaxyFormation2003,
mosterCONSTRAINTSRELATIONSHIPLAR2010,
yangEvolutionGalaxyDarkMatter2012,reddickCONNECTIONGALAXIESDARK2013,
behrooziAVERAGESTARFORMATION2013,mosterGalacticStarFormation2013,
birrerSIMPLEMODELLINKING2014,luEmpiricalModelStar2014,
rodriguez-pueblaConstrainingGalaxyHalo2017,shankarRevisitingBulgeHalo2017,
mosterEmergeEmpiricalModel2018,behrooziUniverseMachineCorrelationGalaxy2019}, 
the modeled stellar mass densities do not 
reach the observational values. This indicates that a simple extrapolation 
of low-z observations is insufficient in explaining the CEERS observations.

All the results presented above highlight the importance of fully 
modeling all the uncertainties when interpreting the observational data. Since each of 
the three sources of uncertainty considered above has a positive effect 
on $\rho_*(>M_*)$, a lower star formation efficiency $\epsilon_*$ is needed 
to explain the observed data when the uncertainty involved is larger.
Conversely, a greater $\epsilon_*$ is necessary to explain the 
observational data if the uncertainty is smaller. Since the exact 
level of uncertainty in each transformation step is not  
known a priori, we provide a general prediction for the relationship 
between the level of uncertainty and the probability to accommodate 
the observational data. For a given observed density, 
$\rho_{*, {\rm obs}}(>M_*)$, we define the probability to observe it as the 
remaining probability mass of having $\rho_*(>M_*)> \rho_{*, {\rm obs}}(>M_*)$:
\begin{equation}
    p_{\rm obs} = 1 - \left. {\rm CDF}[\rho_*(>M_*)] \right\vert _{\rho_*(>M_*)=\rho_{*,{\rm obs}}(>M_*)},
\end{equation}
where ${\rm CDF}[\rho_*(>M_*)]$ is the cumulative distribution function of 
$\rho_*(>M_*)$ derived from $P[\rho_*(>M_*)]$. 
This probability can be used to obtain the level of uncertainties and the star 
formation efficiency required to explain the observation. 

\begin{figure*} \centering
    \includegraphics[width=0.68\textwidth]{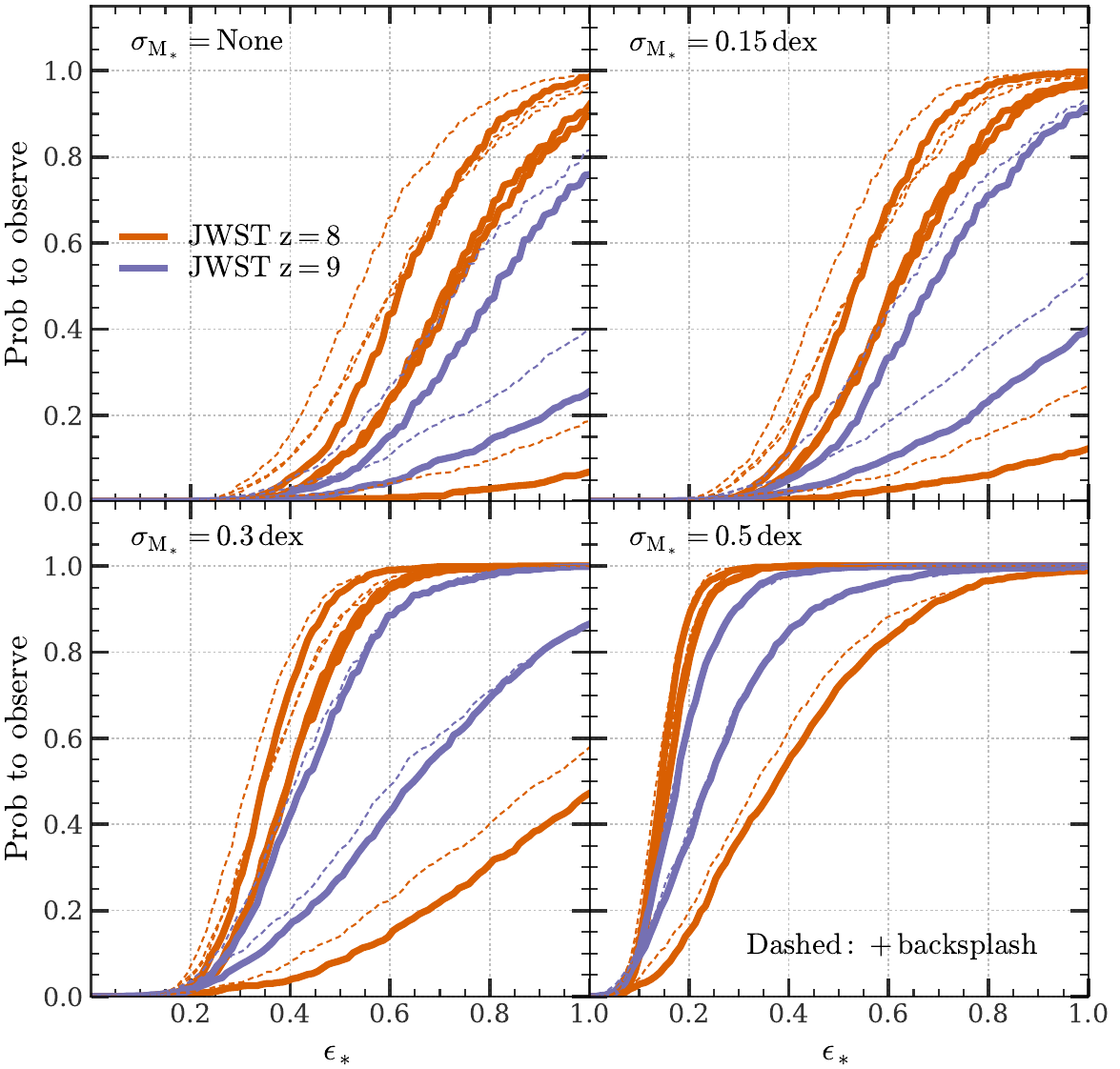}
    \caption{
        The probability of obtaining the cosmic cumulative stellar mass density, 
        $\rho_*(>M_*)$, as measured from the six extremely massive galaxies of 
        the JWST CEERS sample \citep{labbePopulationRedCandidate2023}. The probability 
        is plotted as a function of the star formation efficiency, $\epsilon_*$, 
        used in the conversion from halo mass to stellar mass. Each panel displays 
        the result for a given random error in the estimate of stellar mass, 
        $\sigma_{M_*}$, which is indicated in the \figbf[upper left corner] of that panel. 
        The volume of sub-boxes used to estimate the probability is chosen to be 
        consistent with the JWST CEERS sample at the corresponding redshift, 
        as indicated in the \figbf[first panel]. The \figbf[solid lines] are obtained 
        using simulated halo mass, while the \figbf[dashed lines] are obtained by 
        adding the mass due to backsplash halos. 
        For futher details, refer to \S\ref{ssec:uncertainty-in-jwst}.
    }
    \label{fig:prob_for_jwst_obs}
\end{figure*}

Fig.~\ref{fig:prob_for_jwst_obs} displays the predicted $p_{\rm obs}$ for 
the six data points obtained by \citet{labbePopulationRedCandidate2023} as 
a function of star formation efficiency $\epsilon_*$, taking into account 
different sources of uncertainty. The upper left panel 
shows the case that does not include the random error in the stellar mass estimator. 
Without mass added by backsplash halos, four of the samples have 
$p_{\rm obs} > 10\%$ for $\epsilon_*=0.5$. However, one extreme 
at $z\sim8$ has $p_{\rm obs} < 10\%$ even for $\epsilon_*=1$, suggesting a 
possible tension. With the backsplash effect included, the $p_{\rm obs}$ 
value is three times the original value at $\epsilon_*=1$ for this extreme, 
and it is non-zero at $\epsilon_*=0.5$. A random error with $\sigma_{\rm M_*}$ 
ranging from $0.15\,{\rm dex}$ to $0.3\,{\rm dex}$ on the stellar mass 
estimate increases the probability further. A 
$\sigma_{M_*}$ of 0.5 dex drastically changes the trend of $p_{\rm obs}$; 
even $\epsilon_*=20\%$ is sufficient to move all the observed 
samples to within the $1\sigma$ range ($p_{\rm obs} \geqslant 0.16$).

It is important to note that the probability prediction presented here 
is based solely on empirical modeling of uncertainties in the
$\Lambda$CDM cosmology. Therefore, the results are quite general and 
can be extended to other surveys, regardless of the details of the survey 
strategy, data processing and statistical model. 
However, further work is needed to confirm conjectures presented 
here. Surveys with large areas are useful in suppressing 
cosmic variance and thus in reducing the uncertainty in volume sampling. 
Hydrodynamic simulations and semi-analytical modeling are needed 
to verify or rule out the proposed effect of backsplash, as well as 
to understand the conversion of baryons to stars. Deeper imaging, 
high S/N spectroscopy, and precise stellar population synthesis 
are critical in order to narrow down the posterior parameter 
space in the estimates of redshift and stellar mass estimate. 
All these, together, may eventually resolve the tension 
or provide definitive evidence for the need of new cosmology 
and new physics.

\section{Descendants of High-redshift Massive Galaxies}
\label{sec:descendant}

Within the scope of data and models currently available, it is interesting to 
explore how the observed massive galaxies would evolve over the 
cosmic history and where they end up in the local universe. Answers to 
these questions may provide hints for searching for descendants
of these high-$z$ massive galaxies.

\subsection{Mass Distribution of Descendant Halos}
\label{ssec:descendant-mass-dist}

\begin{figure*} \centering
    \includegraphics[width=\textwidth]{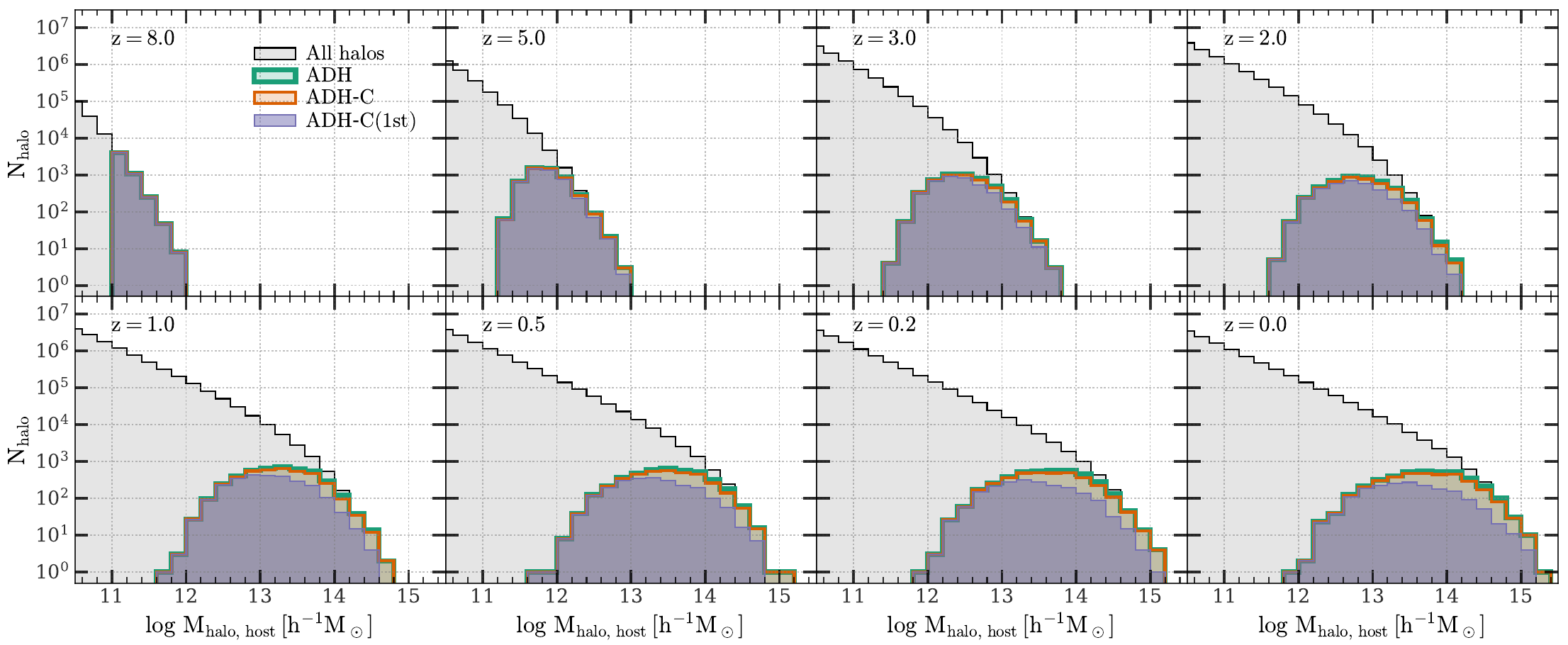}
    \caption{
        Histograms of the halo mass, $M_{\rm halo, host}$, for ancient descendant
        halos (ADH) of a sample of the most massive galaxies, $S$, selected with 
        $M_{\rm halo, host} \geqslant 10^{11} \msun$ at redshift $z_{\rm g}=8$ from ELUCID. In each panel, 
        the \figbf[colored histograms] represent different types of ADH 
        with incremental constraints. The \figbf[green histogram] labeled as ``ADH'' includes 
        any halo that hosts at least one descendant subhalo, whether in 
        central or satellite, of a galaxy in $S$. The \figbf[brown histogram] labeled as ``ADH-C''
        includes any halo whose central subhalo is a descendant of a galaxy in $S$. 
        The \figbf[purple histogram] labeled as ``ADH-C(1st)'' includes any halo whose central subhalo is the 
        first descendant of a galaxy in $S$. For comparison, the histogram for 
        all halos at each snapshot is shown in \figbf[black]. The results suggest 
        that most of the ancient stars formed in the most massive halos at $z_{\rm g} \sim 8$ 
        will eventually reside in the massive halos with $M_{\rm halo, host} \gtrsim 10^{13} \msun$.
        For further details, refer to \S\ref{ssec:descendant-mass-dist}.
    }
    \label{fig:descend_hist}
\end{figure*}

\begin{figure*} \centering
    \includegraphics[width=\textwidth]{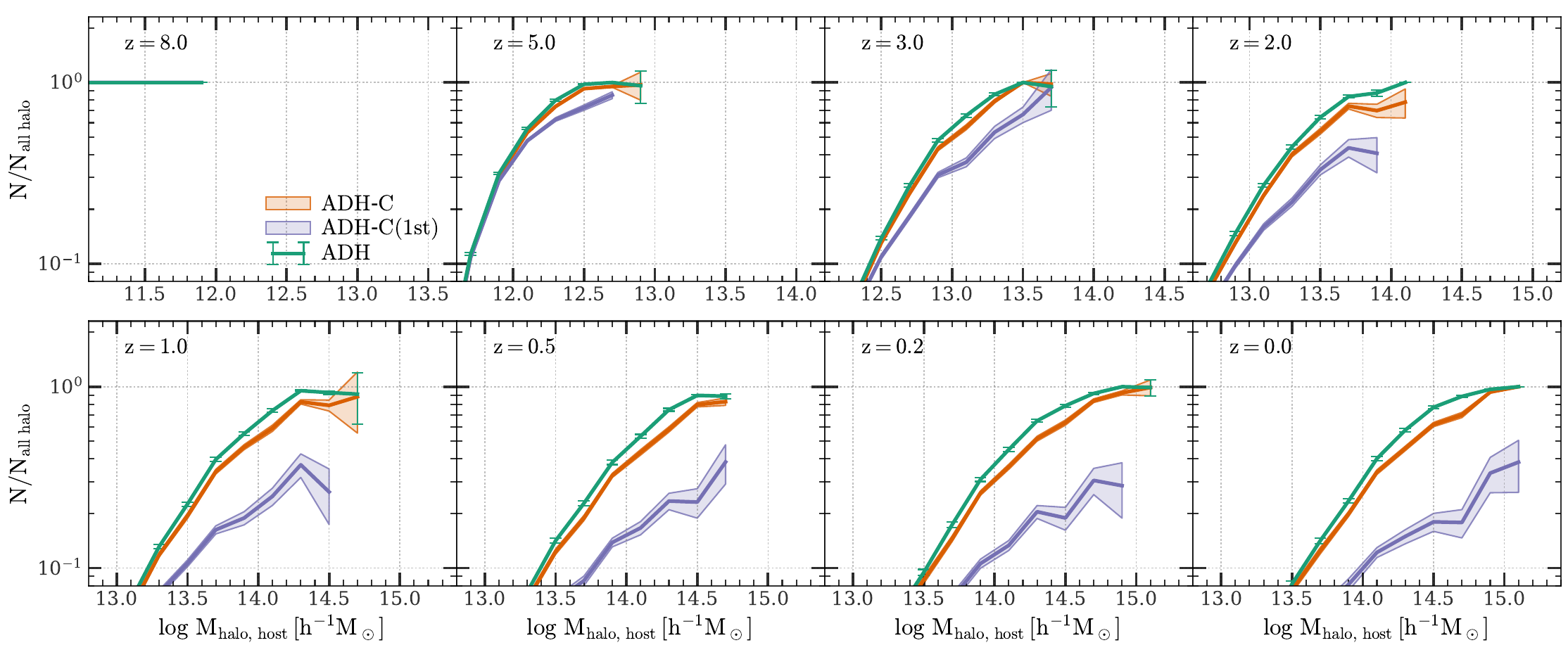}
    \caption{
        The proportions of various types of ancient descendant halos (ADH) of a sample 
        of the most massive galaxies at redshift $z_{\rm g}=8$. These proportions are obtained by calculating 
        the ratios of the colored histograms in each panel of Fig.~\ref{fig:descend_hist} 
        to the black histogram. The mean fraction is represented by the 
        \figbf[solid line], while the \figbf[shaded area] or \figbf[error bar] 
        indicates the standard deviation, which is computed from 100 bootstrapped samples.
        For further details, refer to \S\ref{ssec:descendant-mass-dist}.
    }
    \label{fig:descend_hist_ratio}
\end{figure*}

\begin{figure*} \centering
    \includegraphics[width=\textwidth]{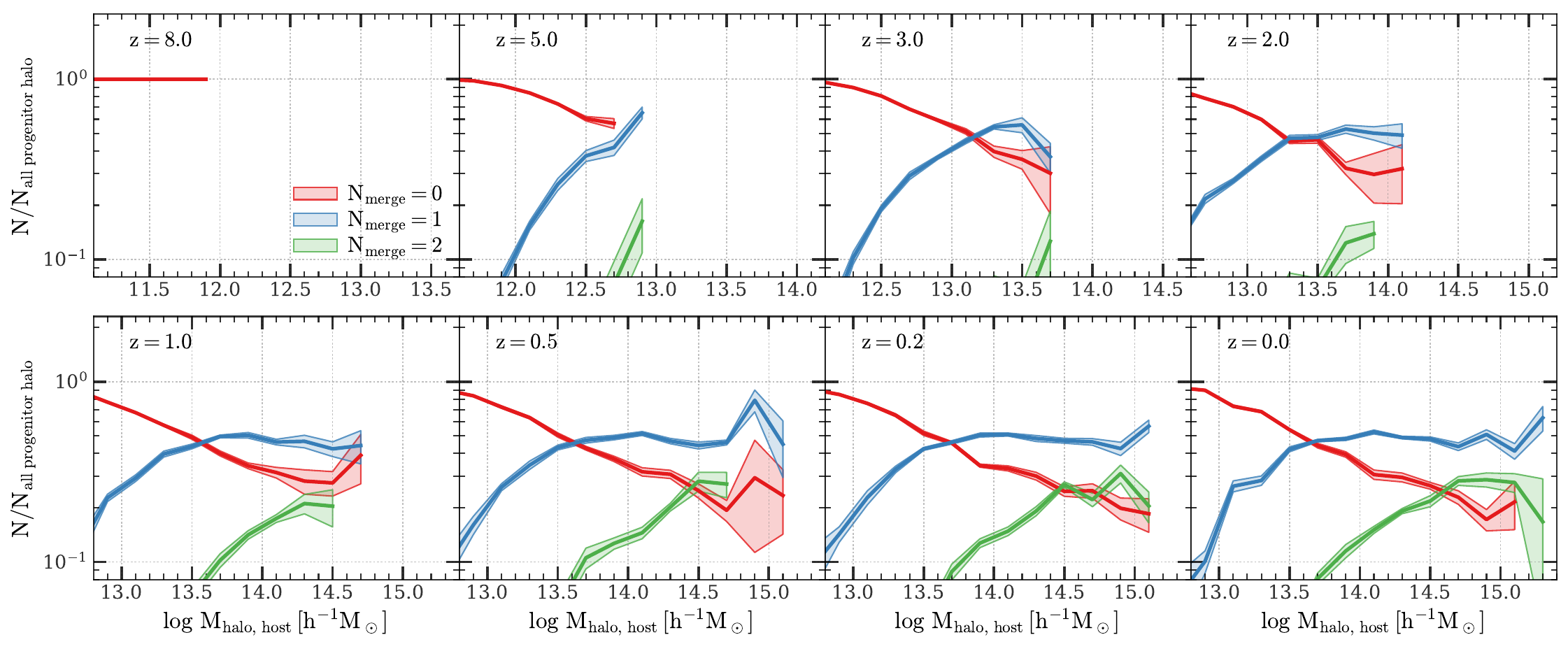}
    \caption{
        The proportion of $z_{\rm g}=8$ massive galaxies that 
        enter the central subhalos of their ancient descendant halos (ADH) at 
        lower $z$, with different numbers of mergers, 
        denoted as $N_{\rm merge}$. 
        The result for each given $N_{\rm merge}$ is represented by \figbf[a different color].
        The halo mass, $M_{\rm halo, host}$, of ADH is labeled on the \figbf[x-axis]. 
        The \figbf[solid line] represents the mean fraction, while the \figbf[shaded area] 
        indicates the standard deviation, which is computed from 100 bootstrapped samples.
        For further details, refer to \S\ref{ssec:descendant-mass-dist}.
    }
    \label{fig:descend_hist_ratio_for_merge}
\end{figure*}

In order to investigate the properties of the descendants of
high-redshift massive galaxies, we use subhalo merger trees in the ELUCID simulation. 
This enables us to establish a connection between the high-redshift galaxies, 
as predicted by our empirical transformations in \S\ref{ssec:method-uncertainties}, 
and low-redshift halos which host the descendants of these galaxies. 
For convenience these low-$z$ halos are referred to as ``ancient descendant halos'', 
or ``ADH'' for short. Note that ADH are defined at a given redshift, $z$, 
for galaxies modeled at a higher redshift, $z_{\rm g}$.     
We study both the spatial and mass distribution of these ADH.
It is important to note that the density field of ELUCID is constrained by 
real observations, as outlined in \S\ref{ssec:data-simulation}. 
Thus, the connection established can also be used to understand the formation 
history of real massive clusters in the local universe back to a time 
when the universe was only approximately 500 million years old.

The green histograms in Fig.~\ref{fig:descend_hist} 
depict the distribution of mass, $M_{\rm halo,host}$, for 
the ADH of high-$z_{\rm g}$ massive galaxies. In this section, 
we use the top-hat mass for halos, since it is easier to infer from 
observation and since we are conducting statistics for a halo as a whole 
\citep[e.g.,][]{yangGalaxyGroupsSDSS2007,yangEvolutionGalaxyDarkMatter2012, 
limGalaxyGroupsLowredshift2017, tinkerSelfCalibratingHaloBasedGalaxy2020,
yanGalaxyClusterMass2020,wangIdentifyingGalaxyGroups2020,hungOpticalObservationalCluster2021}. 
We select high-$z_{\rm g}$ galaxies as the central galaxies in the most 
massive halos with $M_{\rm halo,host} \geqslant 10^{11} \msun$ at $z_{\rm g}=8$ 
in the entire ELUCID volume. From $z=8$ to $z=0$, the descendant halos 
of these massive high-redshift galaxies are found preferentially 
in the massive end of the mass distribution of the total 
population (the black histogram), although the distribution can extend
to masses as low as $\sim 10^{12} \msun$. At the massive end 
of the histogram, almost all halos are ADH of massive galaxies at $z_{\rm g} = 8$. 
The minimum mass of the descendant halo is approximately $10^{11.5}\msun$ 
($10^{12}\msun$) at $z=2$ ($z=0$), while the median is around $10^{12.8} \msun$ 
($13.5 \msun$). However, among the total population of halos 
at $z=0$ with $M_{\rm halo} \sim 10^{13.5} \msun$, only about 
$10\%$ are ADH of massive galaxies at $z_{\rm g}=8$, and the fraction drops 
exponentially towards lower halo masses.

To quantify the precise fraction of high-$z_{\rm g}$ massive galaxies that end up 
in low-$z$ halos of different masses, we present the ratio between 
the ADH mass histogram with the unconditional halo-mass histogram
at different redshift, $z$, in 
Fig.~\ref{fig:descend_hist_ratio} using the same green color as in 
Fig.~\ref{fig:descend_hist}. The shaded area attached to each curve represents 
the standard deviation estimated from 100 bootstrapped samples. A halo with a 
mass of $10^{14} \msun$ ($10^{15} \msun$) at $z=2$ ($z=0$) almost certainly contains 
at least one descendant of the most massive galaxies at $z_{\rm g}=8$, while a 
halo with a mass of $10^{13.5}\msun$ ($10^{14.5}\msun$) has a probability 
of $\gtrsim 70\%$ to do so. For halos with a mass of $10^{13}\msun$ ($10^{13.5}\msun$), 
the probability drops to $\lesssim 30\%$. Note that we do not 
differentiate between centrals and satellites in descendant halos.

The brown histograms labeled as ``ADH-C'' in Fig.~\ref{fig:descend_hist} 
represent the number of ADH that contain the descendant galaxies as their 
central galaxies. The lines in Fig.~\ref{fig:descend_hist_ratio} show the fraction 
of these halos among the total population at the same redshift. The similar values of the 
green (ADH) and brown (ADH-C) histograms/lines suggest 
that most of the high-$z_{\rm g}$ massive galaxies eventually become the dominant 
galaxies or parts of the dominant galaxies of massive 
halos at low-$z$. Due to hierarchical formation of galaxies in the $\Lambda$CDM model, 
some of the high-$z_{\rm g}$ massive galaxies become parts of more dominating 
centrals at low-$z$ while others remain as the dominating centrals. 
To demonstrate this, the purple 
histograms and lines in Fig.~\ref{fig:descend_hist} and Fig.~\ref{fig:descend_hist_ratio}, 
marked as ``ADH-C(1st)'', represent the number and fraction, respectively, 
for low-$z$ halos whose central subhalos are the first descendants of 
the $z_{\rm g}=8$ massive galaxies. 
Here, a subhalo A is considered to be the first
progenitor of another subhalo B in the subsequent snapshot, 
if A has the greatest bound mass among all of B's progenitors.
In this case, we refer to B as the first descendant of A,
only to make clear the special status of A. 
In addition, the chain of first progenitors (i.e., the main branch) 
of B can be determined by recursively identifying the first progenitor 
along the links in the tree towards higher redshift.
Thus, ADH-C(1st) at a given $z$ are halos at $z$ that each has a massive 
progenitor galaxy at $z_{\rm g}$ in the main branch of its central subhalo.
The results indicate that approximately 
$30\% - 40\%$ of the central galaxies in the most massive halos at 
low-$z$ halos are direct descendants of the most massive, 
high-$z_{\rm g}$ galaxies.

The findings presented here provide a compelling guidance regarding the 
sites to identify remnants of the old stellar population 
formed at high-$z$. For instance, since 
central galaxies in the majority of massive clusters with 
$M_{\rm halo} \sim 10^{14} \msun$ ($10^{15} \msun$) at $z=2$ ($z=0$) are 
highly likely to contain ancient stars from $z \sim 8$, targeting the central 
galaxies of these potential ADH with infrared spectroscopic observations 
might reveal the presence of this ancient stellar population. 
This, in turn, would facilitate comparisons 
with stellar populations that formed later to investigate
star formation in high-$z$ galaxies.

In Fig.~\ref{fig:descend_hist_ratio_for_merge}, we partition the $z_{\rm g}=8$ galaxy sample, 
whose descendants are centrals at some lower redshift (indicated by orange 
symbols in Fig.~\ref{fig:descend_hist} and Fig.~\ref{fig:descend_hist_ratio}), 
into subsets based on $N_{\rm merge}$, the number of mergers
they underwent before entering the central galaxies of their ADH. 
Note that here we only include merger events in which the 
subhalo under consideration was the less-massive one in the encountering subhalos.
The fractions of galaxies in these subsets as a function of host halo mass are depicted by lines 
and shadings of different colors for different redshifts below $z=8$. The 
fraction of galaxies with $N_{\rm merge}=0$ consistently declines with 
increasing host halo mass at all redshifts. This is because a host halo with 
greater mass harbors more satellites, thereby increasing the likelihood of 
close encounters among its satellites. The proportion of galaxies with 
$N_{\rm merge}=1$ or $2$ steadily increases with host halo mass at all redshifts. 
At the high-mass extreme, galaxies with $N_{\rm merge}=1$ 
emerge as the dominant population, indicating that most central galaxies in 
the most massive halos at lower redshifts have undergone at least one merger 
event in their history. As redshift decreases, the $N_{\rm merge}=2$ 
cases become increasingly prevalent, eventually outnumbering the 
$N_{\rm merge}=0$ case.

Given that the galaxies we have chosen at $z_{\rm g}=8$ are the most massive ones 
at that time, it is highly probable that
the merger events in which they entered the central galaxies of their ADH
were all major ones. These major mergers can cause angular momentum 
loss, black hole growth, morphology transformation, 
and quenching of galaxies. Our findings thus imply that the central galaxies 
in local massive halos underwent one or two bursts in star formation. 
These outcomes may serve as valuable priors for Bayesian spectral synthesis 
models \citep[e.g.,][]{zhouSDSSIVMaNGAStellar2019,johnsonStellarPopulationInference2021}, 
where one or more burst components can be incorporated into the star formation history.

The significant spread of the $M_{\rm halo, host}$ distribution for ADH
at $z \lesssim 5$ shown in Fig.~\ref{fig:descend_hist}, coupled with the diverse 
merger histories of the descendants of high-$z_{\rm g}$ massive galaxies shown in 
Fig.~\ref{fig:descend_hist_ratio_for_merge}, implies a disruption of the rank order 
for halo/stellar mass during the evolution. Consequently, 
abundance matching between progenitors and descendants using only halo 
and stellar mass may not be accurate, thus presenting a problem for rank-based 
empirical methods to link galaxies between different redshifts 
\citep[e.g.,][]{zhengGalaxyEvolutionHalo2007,behrooziUniverseMachineCorrelationGalaxy2019,
wangRelatingGalaxiesDifferent2023,zhangTrinitySelfConsistentlyModeling2022}. 
Our findings suggest that these methods need to be improved, and that the 
inclusion of secondary halo properties might be needed. We also note that 
some efforts have been made to identify proto-clusters at intermediate redshifts 
($1 \lesssim z \lesssim 3$) and use them to establish a statistical 
link between galaxies at different redshifts
\citep[e.g.,][]{chiangANCIENTLIGHTYOUNG2013,chiangDISCOVERYLARGENUMBER2014,
caiMAppingMostMassive2016,caiMAppingMostMassive2017,wangFindingProtoclustersTrace2021}.
With ongoing and future surveys, these methods can be extended and applied 
to high-$z$ data.
This will open a new avenue to study the evolution of 
the galaxies across the history of the universe. 

\subsection{Assembly History of Individual Clusters}
\label{ssec:descendant-individual}

\begin{figure*} \centering
    \includegraphics[width=\textwidth]{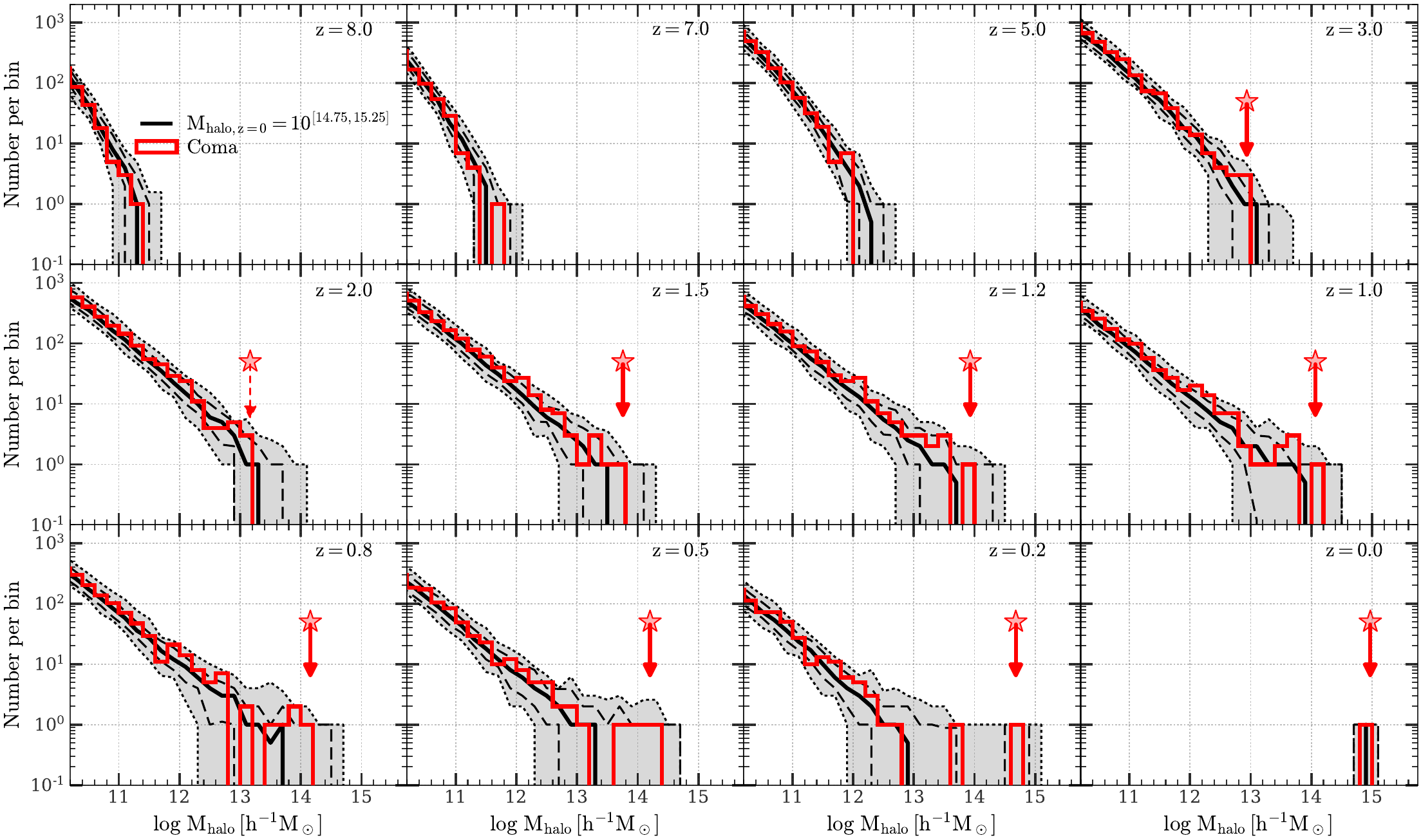}
    \caption{
        The progenitor halo mass functions at different redshifts, shown 
        in \figbf[different panels], of massive halos at $z=0$. 
        The \figbf[black histograms] in each panel correspond to the $z=0$ halos 
        with $10^{14.75} \leqslant \Mhalo/(\msun) \leqslant 10^{15.25}$. 
        The \figbf[solid], \figbf[dashed] and \figbf[dotted lines] indicate 
        the median, $1\sigma$, and $2\sigma$ ranges of the histograms at a given 
        halo mass, respectively. The \figbf[red histogram] represents the Coma cluster, 
        while the \figbf[red arrow] indicates the host halo mass of the first progenitor (main branch) subhalo
        of the central subhalo of Coma, \figbf[solid] if the progenitor is a central subhalo 
        and \figbf[dashed] if it is a satellite. These results reveal an unusual formation 
        history of Coma, whose mass was mostly assembled at $z \lesssim 2$ due to 
        violent merges. For further details, refer to \S\ref{ssec:descendant-individual}.
    }
    \label{fig:des_hmf_m15}
\end{figure*}

\begin{figure*} \centering
    \includegraphics[width=0.9\textwidth]{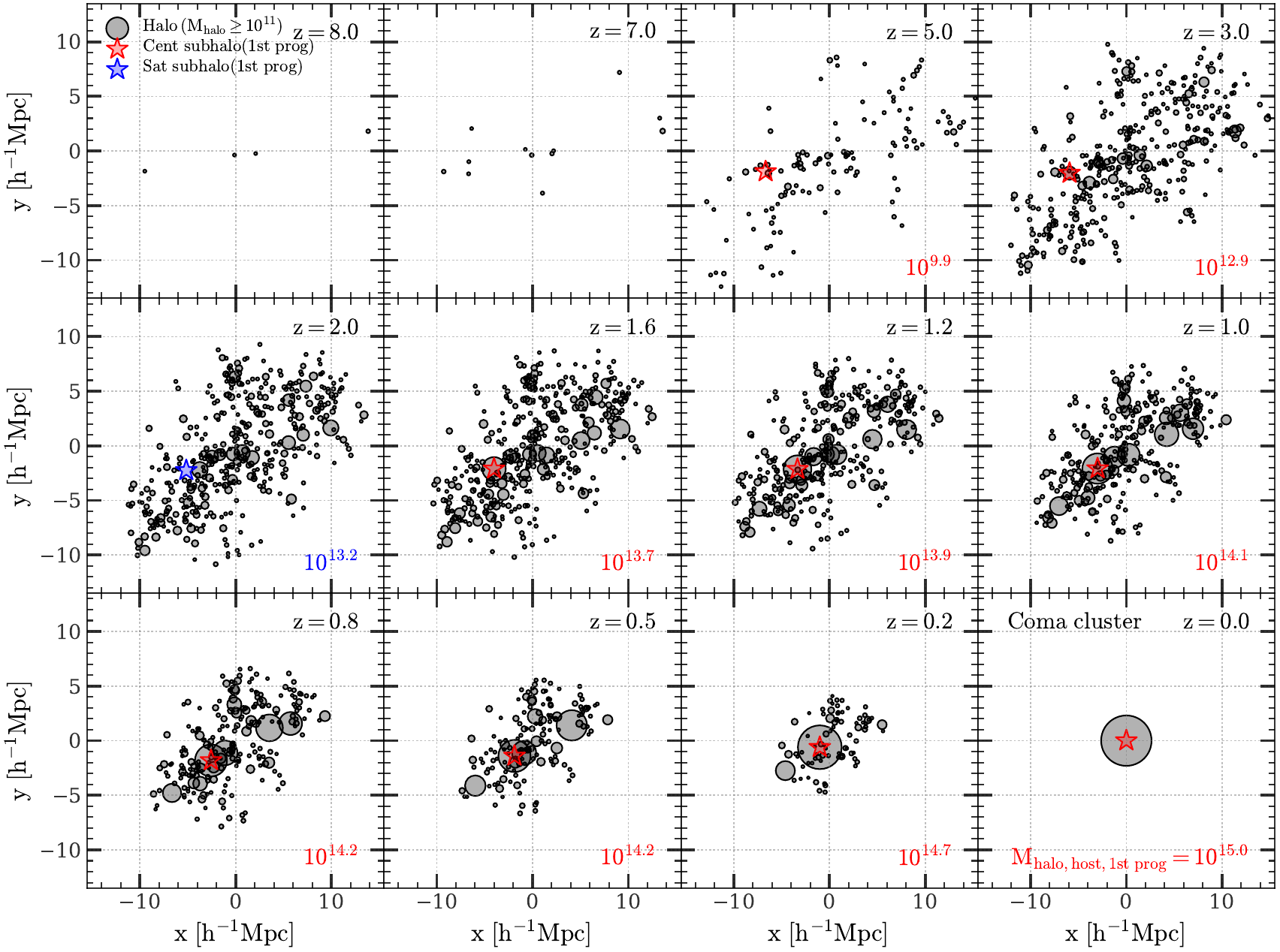}
    \caption{
        The spatial distribution of progenitor halos of the Coma cluster at 
        different redshifts, shown in \figbf[different panels], as simulated by ELUCID. 
        In each panel, each black \figbf[circle] represents one progenitor halo
        with halo mass $M_{\rm halo} \geqslant 10^{11} \msun$. 
        The \figbf[size] of circle is set according to halo mass.
        The \figbf[star] symbol marks the location of the first progenitor (main branch) 
        subhalo of the central subhalo of Coma, shown in
        \figbf[red] if it is a central subhalo and \figbf[blue] if it is a satellite. 
        The host halo mass of the first progenitor subhalo is indicated in the 
        \figbf[lower-right corner] of each panel. 
        For further details, refer to \S\ref{ssec:descendant-individual}.
    }
    \label{fig:scatter-coma}
\end{figure*}

The diversity of the descendant distribution highlights the need to investigate 
the evolution of individual galaxy clusters in the low-$z$ universe. 
In this section, we present the assembly histories of selected halos 
that can be matched to real clusters of galaxies, capitalizing 
on ELUCID with its nature of being constrained by the galaxy 
distribution in the SDSS survey.    

Coma is a massive cluster located in the nearby universe, 
positioned at $({\rm RA}$, ${\rm Dec}$, ${\rm z_{\rm CMB}})$ 
$=(194.95^{\circ}$, $27.98^{\circ}$, $0.0231)$. 
According to \citet{yangGalaxyGroupsSDSS2007,yangEvolutionGalaxyDarkMatter2012}, 
its estimated halo mass is about $9.3\times 10^{14} \msun$. In ELUCID, 
the matched halo is located at $(194.81^{\circ}$, $27.91^{\circ}$, $0.0240)$ 
and has a halo mass of $9.4 \times 10^{14} \msun$. In Fig.~\ref{fig:spatial_scatter}, 
we follow a halo with $M_{\rm halo}$ slightly larger than $10^{11} \msun$ at 
$z = 8$ (top-left panel), which joins the Coma cluster at $z = 0$ (lower-right panel). 
Note that this halo is actually not the most massive one at $z = 5$, $3$ and $1.5$
among all the halos displayed in the same panels. 
Only at lower redshifts, $z = 1.5$ and $0$, does Coma become the most massive 
cluster. This suggests an unusual characteristic of Coma, which formed relatively 
late, perhaps via violent mergers during later epochs.

To understand the evolution of Coma in more detail, Fig.~\ref{fig:des_hmf_m15} 
displays the mass distribution reconstructed by ELUCID for its progenitor halos,
with a red arrow indicating the host halo mass, at each snapshot, 
of the first progenitor subhalo of the central subhalo of Coma at $z=0$. 
For comparison, we depict the median mass distribution, as well as the 1 and 
$2\sigma$ ranges, for progenitors of halos at $z=0$ with 
$10^{14.75} \leq M_{\rm halo}/(\msun) \leq 10^{15.25}$. At $z = 8$, The most massive 
progenitor of Coma has $M_{\rm halo}\sim 2\times 10^{11} \msun$, barely 
touches the high-mass end, as shown in the top-left panel of Fig.~\ref{fig:descend_hist}. 
At $z = 0$, Coma has $M_{\rm halo}\approx 10^{15} \msun$, which is 
at the high-mass end, as demonstrated in the lower-right panel of 
Fig.~\ref{fig:descend_hist}. At $z = 5$, progenitors of Coma are less massive 
than the average for similarly massive halos at $z=0$. At $z = 3$, the first 
progenitor of Coma forms, and at $z = 2$, it merges with another halo and 
becomes a satellite. This suggests that the birthplace of Coma was in a 
crowded environment. At $z \leq 1.5$, the massive end of the distribution 
of Coma's progenitor mass exceeds the average, and the overall amplitude of 
the distribution also goes above the average (as seen from the red and 
black solid histograms). At $z \leq 1$, several massive progenitors form 
and eventually merge with the first progenitor. 
All these suggest that violent mergers at $z \lesssim 3$, particularly 
at $z \leq 1$, played a critical role for Coma to assemble a large  
amount of mass by $z=0$.

In Fig.~\ref{fig:scatter-coma}, we present the projected 2-D spatial positions 
of Coma's progenitors. A star is used to mark the spatial location of the first progenitor 
subhalo, at each snapshot, of the central subhalo of Coma at $z=0$, shown in red if 
the progenitor subhalo is a central and blue if it is a satellite. 
The abundance of progenitors at 
$z \lesssim 5$ indicates the dense environment of Coma, where frequent mergers 
are taking place. This is in agreement with observations, such as those from 
SDSS \citep{abazajianSecondDataRelease2004}, which report a substantial fraction 
of red galaxies in Coma \citep[e.g.,][]{deproprisInfraredLuminosityFunction1998,
eisenhardtMultiapertureUBVRIzJHKPhotometry2007,jenkinsUncoveringNearIRDwarf2007,
adamiVeryDeepSpectroscopy2009,
millerDeepVeryLarge2009,
mahajanStarFormationStarbursts2010,
hammerDeepGALEXObservations2010,mahajanEvolutionDwarfGalaxies2011,
deproprisKbandLuminosityFunctions2017}.
At $z = 0.5$, two massive structures are visible among 
the progenitors, consistent with the observed presence of two massive elliptical 
galaxies near the center of Coma. It is noteworthy that overall mass distribution 
of Coma is not spherically symmetric and that it is surrounded by filamentary 
structures that contribute to its continuous mass assembly. Future X-ray 
observations of these filaments may provide additional constraints on the 
environment and assembly history of the Coma cluster. 

\begin{figure*} \centering
    \includegraphics[width=\textwidth]{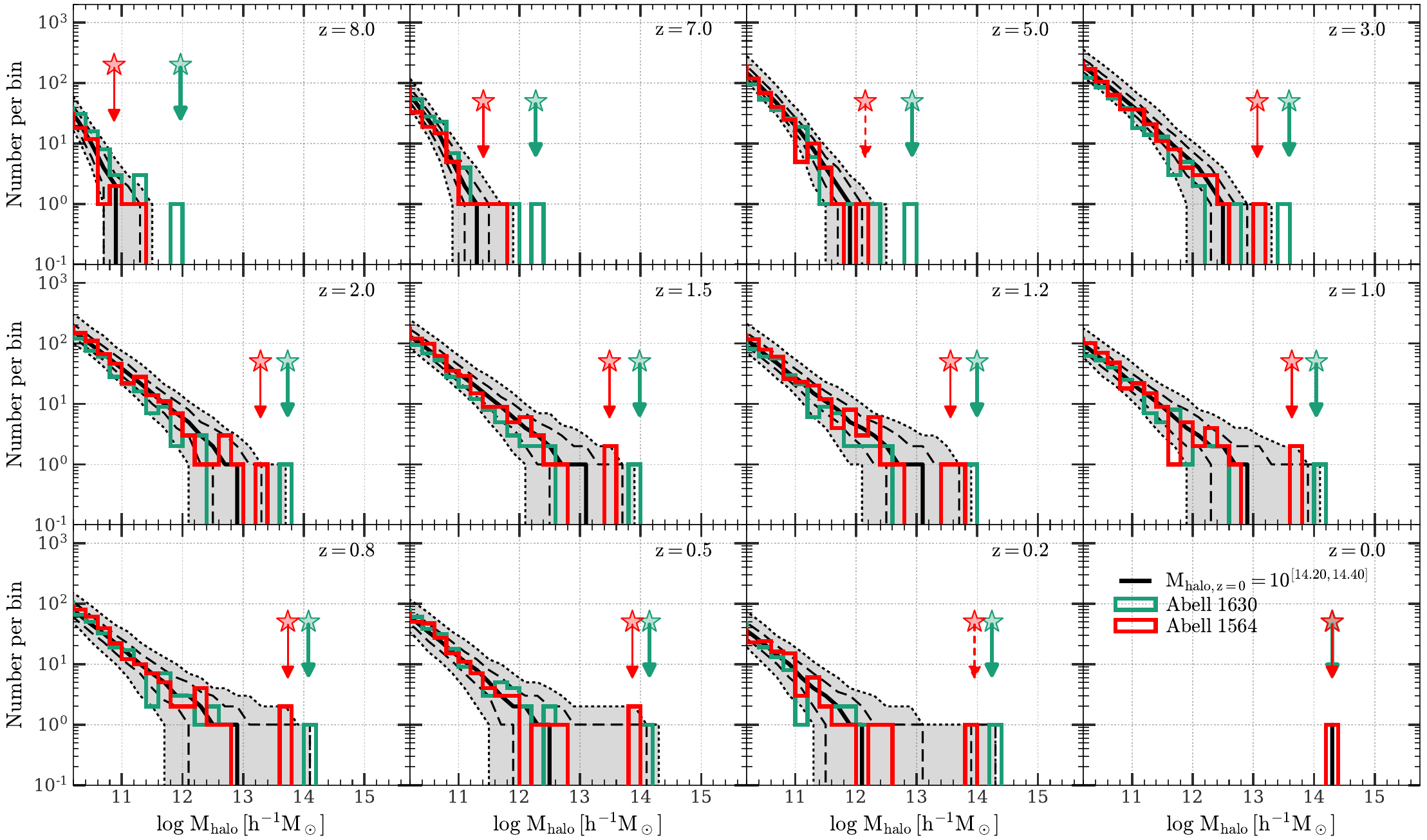}
    \caption{
        The same as Fig.~\ref{fig:des_hmf_m15} but for less massive $z=0$ halos 
        with $10^{14.2} \leqslant \Mhalo/(\msun) \leqslant 10^{14.4}$. \figbf[Green]
        and \figbf[red] histograms/arrows are for two $z=0$ clusters, Abell 1630 and 
        Abell 1564, respectively. These results show two extremes of 
        cluster formation, through equal-mass merger event (Abell 1564)
        or continuous accretion (Abell 1630).
    }
    \label{fig:des_hmf_m143}
\end{figure*}

\begin{figure*} \centering
    \includegraphics[width=0.9\textwidth]{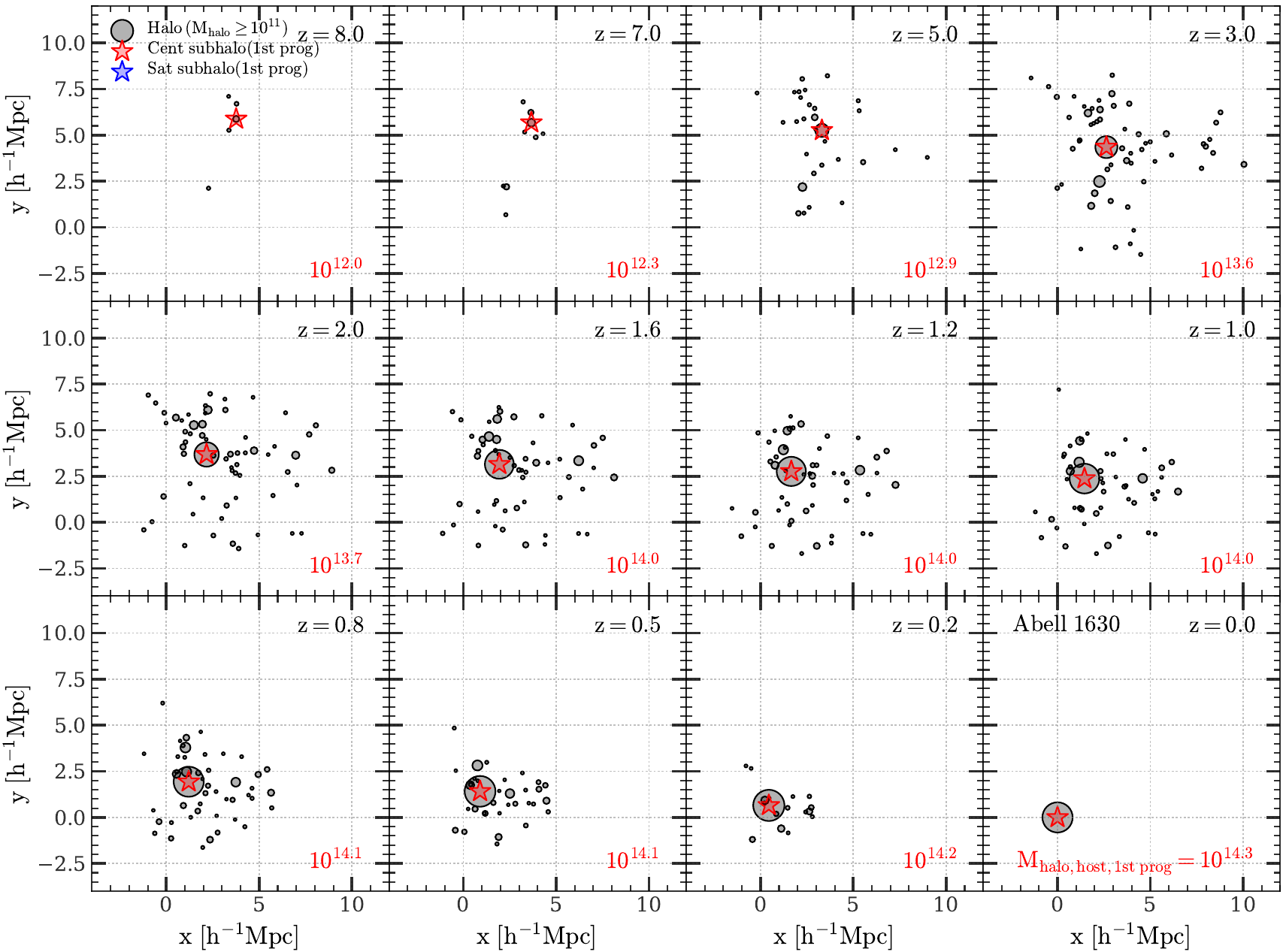}
    \caption{
        The same as Fig.~\ref{fig:scatter-coma} but for Abell 1630.
    }
    \label{fig:scatter-a1630}
\end{figure*}

According to the extended Press-Schechter formalism 
\citep{laceyMergerRatesHierarchical1993},
the progenitor mass distribution is expected to depend on the mass of 
descendant halos. Fig.~\ref{fig:des_hmf_m143} depicts  
the progenitor mass distributions at various redshift for present-day 
halos in the mass range $10^{14.2} \leq M_{\rm halo}/(\msun) \leq 10^{14.4}$.
The median, $1 \sigma$, and $2 \sigma$ ranges are displayed by solid,  
dashed and dotted lines, respectively. In addition, we also show 
the progenitor distributions for two $z=0$ halos matched 
with Abell 1630 and Abell 1564.

Abell 1630 is a galaxy cluster located at $({\rm RA}$, ${\rm Dec}$, ${\rm z_{\rm CMB}})$ 
$ = (192.93^{\circ}$, $4.56^{\circ}$, $0.064)$, and its estimated halo mass 
is $1.78 \times 10^{14} \msun$ by \citet{yangGalaxyGroupsSDSS2007,
yangEvolutionGalaxyDarkMatter2012}. In the ELUCID simulation, its position is 
simulated at $(193.18^{\circ}, 4.42^{\circ}, 0.064)$, and its halo mass is 
$2.01 \times 10^{14} \msun$. Abell 1630 is quite unique in that its first
progenitor significantly dominates the progenitor population. 
The mass gap between the first and other progenitors is evident from 
$z=8$ and continues throughout its history until $z=0$. This suggests that 
Abell 1630 has assembled most of its mass through continuous accretion 
or relatively minor mergers, which is very different from Coma, where late-time 
major mergers dominate the mass assembly. This cluster 
has a massive progenitor at $z=8$, which lies well outside the 
$2\sigma$ range of the mass distribution and is even more massive than 
the most massive progenitor of Coma at the same redshift. 
These suggest that Abell 1630 should have a dominating member at $z=0$ 
near the center. Indeed, a luminous red galaxy with an 
absolute magnitude of $M_{\rm r}^{0.1}=-22.17$ and a color index of 
$(g-r)^{0.1} = 0.93$ has been identified in this cluster, 
consistent with the expectation of ELUCID.

The spatial locations of the progenitors of Abell 1630 at different snapshots 
are displayed in Fig.~\ref{fig:scatter-a1630}. A comparison with the corresponding 
plot for Coma in Fig.~\ref{fig:scatter-coma} reveals that Abell 1630 formed in 
an environment with only a moderate matter density, which may be the cause 
of its assembly history distinctive from that of Coma. 

\begin{figure*} \centering
    \includegraphics[width=0.9\textwidth]{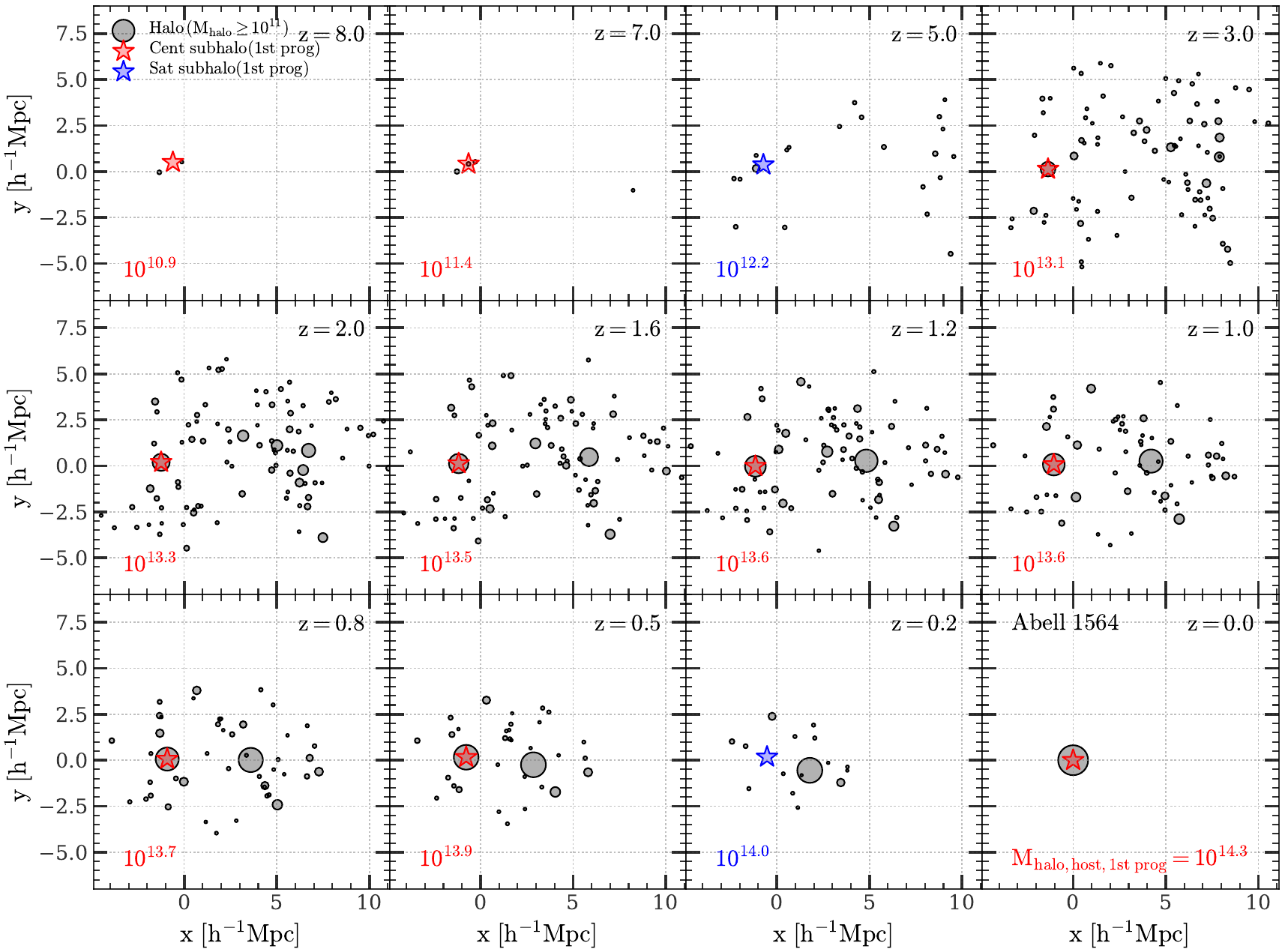}
    \caption{
        The same as Fig.~\ref{fig:scatter-coma} but for Abell 1564.
    }
    \label{fig:scatter-a1564}
\end{figure*}

Abell 1564, the final example, is observed at $(188.74^{\circ}$, $1.84^{\circ}$, $0.0780)$. 
Its estimated halo mass is $2.87\times 10^{14} \msun$, and it is reconstructed by 
ELUCID at $(188.96^{\circ}$, $2.02^{\circ}$, $0.0784)$ with a halo mass of 
$1.99 \times 10^{14} \msun$. The progenitor mass and spatial distributions 
are shown in Fig.~\ref{fig:des_hmf_m143} (the red histograms)
and in Fig.~\ref{fig:scatter-a1564}, respectively.

The assembly history of Abell 1564 falls between the two extremes shown above
in that it has gone through only one major merger in its entire history. 
The two merging progenitors at $z=0.5$ have nearly equal mass and, as a result, 
the first progenitor of Abell 1564 became a satellite subhalo during 
a short period of time at $z\sim 0.2$ after the merger. This late-time major merger 
implies that the cluster at $z=0$ may not yet be fully virialized. 
The non-spherical light distribution of Abell 1564 in real observations 
provides supports to our conclusion.

\begin{figure} \centering
    \includegraphics[width=0.9\columnwidth]{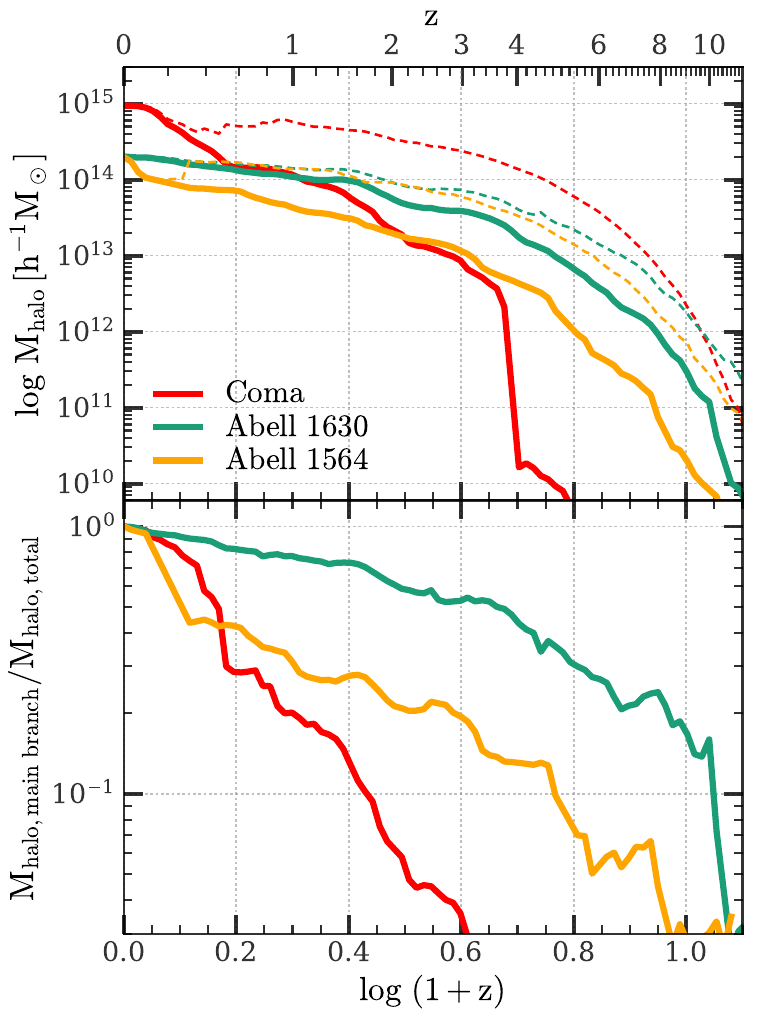}
    \caption{
        Mass assembly histories of Coma (\figbf[red]), Abell 1630 (\figbf[green])
        and Abell 1564 (\figbf[orange]). In the \figbf[upper panel], \figbf[solid lines]  
        show $M_{\rm halo,main\ branch}$, the halo mass of main branch progenitor,
        and \figbf[dashed lines] show $M_{\rm halo,total}$, the total mass contained in
        all progenitor halos with $M_{\rm halo} \geqslant 10^{10}\msun$.
        The ratio of these two masses is shown in the \figbf[lower panel].
    }
    \label{fig:des_mah}
\end{figure}

The three examples presented here illustrate three distinct assembly modes 
of massive clusters: one with multiple major mergers like Coma, one with only one 
major merger like Abell 1564, and one with continuous accretion (or minor mergers)
like Abell 1630. 
To illustrate the differences between the modes, we present the mass assembly 
histories of the three clusters as a function of redshift in Fig.~\ref{fig:des_mah} 
for $M_{\rm halo,main\ branch}$, which is the mass of the main branch progenitor, 
$M_{\rm halo, total}$, which is the total mass contained in all progenitor halos 
with $M_{\rm halo} \geqslant 10^{10}\msun$, and the ratio between them. Among 
the three clusters, Coma is the most massive at $z=0$, but its main progenitor 
is the least massive at $z \gtrsim 3$. Three faster growth stages at $z \sim 5$, 
$2$ and $0.5$ respectively, are clearly seen in the growth of Coma's main branch, 
indicating violent merger events in its history. From Fig.~\ref{fig:scatter-coma}, 
it is evident that these three fast accretion stages result from the richness 
of Coma's progenitor halos and their asymmetric spatial distribution. The other 
extreme, Abell 1630, shows continuous growth of its main branch. It was massive 
enough to host a bright galaxy at $z \gtrsim 8$, but its main-branch mass assembly 
history is smooth and slow, culminating in a final halo less massive than Coma. 
The in-between case, Abell 1564, shows only one main-branch mass jump at $z \sim 0.5$, 
which results from the major merger event seen in Fig.~\ref{fig:scatter-a1564}. 
During most of its history, Abell 1564 has a main-branch-to-total ratio 
lying between the two extremes, Coma and Abell 1630. Note that in some cases during 
mergers, the curves of total mass even decrease. This is an artificial 
effect, due to the fact that some particles linked by the FoF algorithm are not 
enclosed in the "tophat" filter that defines the halo mass. The diversity of the 
formation mode, as represented by the main branch growth rate and the number of 
merger events, may be responsible for the significant variation observed in the 
descendant mass distribution in \S\ref{ssec:descendant-mass-dist}. Such variation 
must be carefully taken into account when constructing models to link galaxies 
over a wide range of redshifts.

\section{Summary and Discussion}
\label{sec:summary}

In this paper, we have developed a simple model to empirically map halos 
from N-body simulations to galaxies 
(\S\ref{sec:data-and-method}). The model consists of four stages, including 
volume sampling, halo mass estimation, stellar mass estimation, and 
statistical analysis. We have incorporated uncertainties from various sources 
in a general manner and integrated them into these stages. The forward 
nature of our model and our detailed treatment of errors have led to 
the following conclusions that have significantly implications 
for the interpretation of recent JWST data:
\benum
\item We have considered uncertainties such as baryon mass enhancement 
by backsplash halos, cosmic variance, and systematic and random errors in stellar 
mass estimation. Due to Eddington bias amplification at the high-stellar-mass end, 
each of these uncertainties can increase the galaxy number density or 
cosmic stellar mass density by an order of magnitude relative to 
pure theoretical expectation. Consequently, the combination of these 
uncertainties can boost the observed density by orders of magnitude 
(\S\ref{ssec:uncertainties-expected}).
\item By applying our model to the sample of extremely massive galaxies at 
$z\sim 7-10$ discovered by \citet{labbePopulationRedCandidate2023} using 
JWST CEERS, we have found that the observed high stellar mass density does 
not present a significant tension with the standard $\Lambda$CDM 
paradigm of structure formation. A reasonable star formation efficiency of 
$\epsilon_*=0.5$ is sufficient to reproduce the observational data 
when cosmic variance is taken into account. The incorporation of backsplash mass 
enhancement further reduces the tension to approximately $1\sigma$
(\S\ref{ssec:uncertainty-in-jwst}).
\eenum

To test the possibility of incorporating the newly discovered extremely massive 
high-$z$ galaxies into the full picture of galaxy formation, it is essential 
to statistically link them with galaxies at lower $z$. This can be achieved 
through observational or theoretical means. In this study, we use a constrained 
simulation, ELUCID, to theoretically predict the descendant halo properties of 
high-$z$ galaxies and to motivate future observations. Our main 
conclusions are as follows:
\benum
\item High-redshift galaxies with masses $\gtrsim 10^{11} \msun$ at $z \sim 8$ 
are expected to fall into the most massive halos at lower-$z$, with a significant 
portion ending up in halos with $M_{\rm halo} \gtrsim 10^{13} \msun$ at $z=0$. 
The central galaxies of galaxy clusters at the high-mass end almost certainly 
contain the stars from massive galaxies at $z\sim 8$. Roughly 40\% of the 
centrals in the most massive clusters at $z = 0$ are the primary descendants of 
these ancient galaxies. Most of the massive galaxies at high-$z$
experienced one major merger before ending up as central galaxies of 
low-redshift massive clusters (\S\ref{ssec:descendant-mass-dist}).

\item By matching the reconstructed $z=0$ halos with observations, 
we find that there is a diversity of assembly histories for massive halos. 
The reconstruction suggests that Coma is an outlier that deviates from the 
regular assembly history of the most massive halos. Its primary structure formed 
relatively late at $z\sim 3$ and then grew rapidly through violent major mergers. 
Another two less massive clusters, Abell 1630 and 1564, exhibit two distinct 
assembly histories, one with continuous accretion and one with an equal-mass 
major merger (\S\ref{ssec:descendant-individual}).
\eenum

While a $\epsilon_* \sim 0.5$ falls within the cosmological upper limit, 
it may still pose a challenge to current galaxy formation theory, given 
that inferred values of $\epsilon_*$ for low-$z$ galaxies are 
generally lower than $0.2$ \citep[e.g.,][]{guoFurtherEvidencePopulation2019,
boylan-kolchinStressTestingLambda2023}. 
Recent measurements of halo mass with weak lensing techniques 
have indicated a $\epsilon_* \sim 0.6$ for massive local star-forming galaxies 
\citep[][]{zhangMassiveStarFormingGalaxies2022}. 
However, these galaxies are very different from the high-$z$ 
massive galaxies in the time scale of star formation. 
\citet{finkelsteinCEERSKeyPaper2023} 
have suggested some possible solutions to the underestimation of UV 
luminosity in semianalytical models and hydrodynamic simulations, 
including modifications to star formation laws, improvements in 
resolving halo merger trees and cool gas clouds, and variations in the IMF.
However, it remains unclear whether or not $\epsilon_* \sim 0.5$ 
can be realized in models of galaxy formation.

Our analysis shows that backsplash halos may serve as an external source of 
baryons for galaxy formation. Evidence for this type of interaction is 
provided by the Bullet cluster in the local universe, where baryons 
become detached from dark matter \citep[e.g.,][]{cloweWeakLensingMass2004, 
markevitchDirectConstraintsDark2004}. 
High baryon fraction has also been observed in dark-matter-deficient 
galaxies (DMDGs) in the local universe. For instance, 
\citet{guoFurtherEvidencePopulation2019} identified 19 such galaxies 
and found that their baryon fraction can be significantly larger 
than that expected from their halo masses scaled with the  
universal baryon fraction. Further analyses using hydrodynamic simulations 
suggested a number of possibilities to form DMDGs. For example, 
using both particle-based and mesh-based high-resolution simulations,
\citet{shinDarkMatterDeficient2020} and \citet{leeDarkMatterDeficient2021}
found that a high-velocity ($\sim 300\,{\rm km/s}$) collision of two gas-rich, 
dwarf-sized galaxies can separate dark matter from warm gas in the disk, 
generate shock compression, and trigger the formation of star clusters 
and eventually the formation of a DMDG. They also found that 
IllustrisTNG100-1, a cosmological simulation with relatively low resolution, 
cannot reproduce the collision-induced DMDG. 
\citet{jacksonDarkmatterdeficientDwarfGalaxies2021} employed high-resolution 
hydrodynamic simulations and suggested another formation channel for DMDGs 
via sustained stripping by nearby massive companions. However, 
these observational and theoretical studies primarily focused on the 
formation of dwarf galaxies with $M_* \lesssim 10^{9}\Msun$. 
Further investigations using simulations of high resolution and 
large volume are needed to determine whether or not similar 
effects can also be produced for more massive systems at high redshift.

It is critical to conduct subsequent spectroscopic observations with NIRSpec 
and MIRI on JWST to confirm the redshift and mass estimates for the candidates 
obtained from photometry data. Already, a small sample of spectroscopically 
confirmed galaxies has been presented by \citet{arrabalharoSpectroscopicConfirmationCEERS2023} 
and \citet{harikaneComprehensiveStudyGalaxies2023}. It is worth noting that a small 
fraction of the high-$z$ candidates identified earlier 
are actually found to be interlopers from low-redshift, 
so that the tension between the observational data and theoretical 
expectations is alleviated. In any case, the results presented 
here represent generic predictions of the current $\Lambda$CDM 
model, and can be used to guide the interpretation of high-$z$
data expected from JWST.

The constrained simulation, ELUCID, is used in our analysis to track the evolution
histories of galaxy clusters in the real Universe. This approach mitigates the 
uncertainties inherent in studies relying solely on statistical analysis. 
However, due to the complexity and non-linearity involved in cluster formation, 
as well as the inevitable uncertainties in observations, the solution to 
cluster assembly history cannot be unique, but rather forms an ensemble that 
follows a posterior distribution surrounding the optimal point identified by ELUCID's 
reconstruction method. Sampling from this high-dimensional posterior space and 
making predictions based on the obtained sample require a precise formulation of the 
posterior distribution, which is currently unknown, and a substantial amount of 
computational resources, which are currently unfeasible. 
The implementation of GPU-based computation acceleration and the adjoint method for 
memory conservation, as suggested by \citet{liDifferentiableCosmologicalSimulation2022},
offer a promising solution to the existing hardware limitations, 
warranting exploration in future investigations.

\section*{Acknowledgements}

YC is funded by the China Postdoctoral Science Foundation (grant No. 2022TQ0329)
and National Science Foundation of China (grant No. 12192224). 
The authors would like to express their gratitude to the Tsinghua Astrophysics 
High-Performance Computing platform at Tsinghua University and the 
Supercomputer Center of the University of Science and Technology of China for 
providing the necessary computational and data storage resources that have 
significantly contributed to the research results presented in this paper.
The authors thank ELUCID collaboration for making their data products publicly available.
YC expresses gratitude to Huiyuan Wang, Yu Rong, Enci Wang, Boseong Cho, 
and Ethan Nadler for their valuable insights and discussions.

\section*{Data availability}

The computations and presentations in this paper are supported by various software 
tools, including the HPC toolkits 
\softwarenamestyle[Hipp] \citep{2023ascl.soft01030C}
and \softwarenamestyle[PyHipp]\footnote{\url{https://github.com/ChenYangyao/pyhipp}},
numerical libraries 
\softwarenamestyle[NumPy] \citep{harrisArrayProgrammingNumPy2020}, 
\softwarenamestyle[Astropy] \citep{
robitailleAstropyCommunityPython2013,
astropycollaborationAstropyProjectBuilding2018,
astropycollaborationAstropyProjectSustaining2022}, 
and \softwarenamestyle[SciPy] \citep{virtanenSciPyFundamentalAlgorithms2020}, 
as well as the graphical library 
\softwarenamestyle[Matplotlib] \citep{hunterMatplotlib2DGraphics2007}. 
The ELUCID project\footnote{\url{https://www.elucid-project.com/}}
provides access to the data used in this study. Additionally, an 
online cosmic variance calculator and its Python API will be accessible from 
the Software page of the ELUCID website.

\bibliographystyle{mnras}
\bibliography{references}

\bsp	
\label{lastpage}
\end{document}